\documentclass[%
 reprint,
superscriptaddress,
 amsmath,amssymb,
 aps,
prb,
floatfix,
showkeys
]{revtex4-2}

\usepackage{graphicx}
\usepackage{dcolumn}
\usepackage{gensymb}
\usepackage{bm}
\usepackage{color}
\usepackage{xcolor}

\begin{document}


\title{From biting to engulfment: Target mechanics determines modes of phagocytosis through curvature--actin coupling}


\author{Shubhadeep Sadhukhan}
\email{shubhadeep.sadhukhan@weizmann.ac.il}
 \affiliation{Department of Chemical and Biological Physics, Weizmann Institute of Science, Rehovot, Israel}
 
\author{Caitlin E. Cornell}
\affiliation{
Department of Bioengineering, University of California Berkeley; Berkeley, CA USA
}
\author{Mansehaj Kaur Sandhu}
\affiliation{%
Department of Molecular Biology and Biochemistry, Simon Fraser University, Burnaby BC, Canada
}%
\author{Marta Batet Palau}
\affiliation{%
Centre for Genomic Regulation (CRG), The Barcelona Institute of Science and Technology, 08003 Barcelona, Spain
}%
\affiliation{%
Universitat Pompeu Fabra (UPF), Barcelona, Spain
}%
\author{Youri Peeters}
\affiliation{%
Department of Cell Biology and Immunology, Wageningen University and Research, Wageningen, the Netherlands
}%
\author{Stijn Hanssen}
\affiliation{%
Department of Cell Biology and Immunology, Wageningen University and Research, Wageningen, the Netherlands
}%
\author{Samo Peni\v{c}}
\affiliation{Laboratory of Physics, Faculty of Electrical Engineering, University of Ljubljana, Ljubljana, Slovenia}
\author{Ale\v{s} Igli\v{c}}
\affiliation{Laboratory of Physics, Faculty of Electrical Engineering, University of Ljubljana, Ljubljana, Slovenia}
\author{Daan Vorselen}
\affiliation{%
Department of Cell Biology and Immunology, Wageningen University and Research, Wageningen, the Netherlands
}%
\author{Daniel A. Fletcher}
\affiliation{
Department of Bioengineering, University of California Berkeley; Berkeley, CA USA
}
\affiliation{University of California Berkeley/University of California San Francisco Graduate Group in Bioengineering, CA USA}
\affiliation{Division of Biological Systems and Engineering, Lawrence Berkeley National Laboratory; Berkeley CA USA}
\affiliation{Chan Zuckerberg Biohub; San Francisco CA USA}
\author{Valentin Jaumouill\'{e}} 
\affiliation{%
Department of Molecular Biology and Biochemistry, Simon Fraser University, Burnaby BC, Canada
}%
\author{Verena Ruprecht}
\affiliation{%
Centre for Genomic Regulation (CRG), The Barcelona Institute of Science and Technology, 08003 Barcelona, Spain
}%
\affiliation{ University of Innsbruck, 6020 Innsbruck, Austria}
\author{Nir S. Gov}
\email{nir.gov@weizmann.ac.il}
 \affiliation{Department of Chemical and Biological Physics, Weizmann Institute of Science, Rehovot, Israel}
\affiliation{Department of Physiology, Development and Neuroscience, Downing Site, University of Cambridge, Cambridge, UK}

\date{\today}

\begin{abstract}

Phagocytosis is a core innate immune process that clears targets spanning a wide range of mechanical properties, yet the role of target mechanics in recognition and engulfment remains unclear. Here, we combine theoretical modeling and experiments to reveal how target stiffness governs distinct modes of phagocyte–target interaction. We develop a membrane-based simulation framework in which both the engulfing cell and its target are deformable and undergo large shape changes, while actin-driven protrusions are regulated by curvature-sensitive membrane complexes. The model predicts three mechanical regimes with increasing target stiffness: (i) biting (trogocytosis), where part of the target is extracted; (ii) pushing, where the target is displaced rather than engulfed; and (iii) complete engulfment. We validate these predictions in epithelial clearance of apoptotic targets in vivo and macrophage engulfment of Giant Unilamellar Vesicles (GUVs) and lymphoma cells. Together, our results identify target mechanics as a key regulator of clearance and cell–cell interactions.
\end{abstract}

\maketitle
\textbf{Significance statement}---{Phagocytosis is essential for immune defence, yet the physical principles governing engulfment of deformable targets remain poorly understood. Most theoretical models assume rigid particles, appropriate for phagocytosis of bacteria or fungi. When phagocytes engage dying cells or antibody-opsonised cancer cells, these targets undergo substantial shape changes during phagocytosis. We develop a theoretical model to simulate cell--cell interactions, enabling a mechanistic exploration of phagocytosis of soft targets. We utilize a model in which cytoskeletal protrusive activity is guided by curvature-sensitive membrane complexes, and show that target membrane rigidity dictates whether targets are fully engulfed, pushed away, or partially bitten. These mechanically driven dynamic regimes are validated experimentally using artificial elastic beads, GUVs, and lymphoma cells, both in vivo and  in vitro.}

\section*{Introduction}\label{sec:intro}
Phagocytosis is essential for immune defense and tissue homeostasis, enabling cells to eliminate pathogens, apoptotic cells, and cancer cells through adhesion and engulfment \cite{flannagan2012cell,uribe2020phagocytosis}. Although target recognition is primarily mediated by molecular signals, how target mechanics regulates phagocyte--target interactions remains poorly understood. Here, we combine theory and experiment to show that target stiffness is a key determinant of phagocytic behaviour.

Phagocytosis relies on dynamic actin recruitment to drive membrane protrusions that progressively wrap around the target \cite{swanson2008shaping,mylvaganam2021cytoskeleton,PANAH2026651}. However, how the actin recruitment and the resulting cytoskeletal force is coordinated with the membrane dynamics during this process is not well understood \cite{vorselen2020mechanical,jaumouille2020physical}. We have recently shown that the coordinated recruitment of actin during phagocytosis of rigid objects can be explained very well by a theoretical model which includes a coupling between curved membrane protein complexes (CMC) that recruit and nucleate actin polymerization \cite{sadhu2023theoretical}. This model demonstrated that the presence of passive CMC can enhance the engulfment process, even in the absence of active forces. When active protrusive forces (which represent the protrusive forces due to actin polymerization) that are recruited by the CMC were considered, the model explained the more robust engulfment, at lower adhesion energies and less sensitive to the object’s shape \cite{champion2006role} observed in cells \cite{sadhu2023theoretical}. By including the effects of active cytoskeletal forces, which are guided by the curvature coupling, this model goes beyond the theoretical descriptions of passive engulfment driven purely by adhesive forces \cite{tollis2010zipper,vacha2011receptor,agudo2015critical,yi2016incorporation}. Curved membrane proteins that recruit actin polymerization have been found experimentally to be associated with the leading edge of cellular protrusions \cite{scita2008irsp53,linkner2014inverse}, which are involved in cellular adhesion, spreading and motility \cite{begemann2019mechanochemical,pipathsouk2021wave,wu2025wave}. The theoretical model demonstrated that this curvature-actin activity coupling can explain many cellular shape and migration dynamics \cite{sadhu2023minimal,sadhu2024minimal,Sadhukhan2025}.

Motivated by these results, we explore here the interactions between a cell and a non-rigid object. We developed a minimal membrane-based model in which actin-driven protrusions emerge through curvature-sensitive membrane complexes and interact with deformable targets. We extend our model to allow for two vesicle-like surfaces to evolve and interact. For two symmetric vesicles containing passive CMC the model predicts a spontaneous symmetry-breaking transition, where one vesicle engulfs the other. For active CMC (describing the effects of actin polymerization in cells) interacting with a soft passive vesicle, the model predicts different dynamic regimes which depend on the rigidity of the target: with increasing rigidity, the active vesicle transitions from “biting”, to “pushing” and eventually “engulfing” the target. Despite its simplicity, the model predicts three distinct modes of phagocyte–target interaction that arise as a function of target stiffness: trogocytosis-like biting of soft targets, pushing of deformable targets without engulfment, and complete engulfment of stiff targets. These behaviors emerge from mechanical interactions alone, without invoking stiffness-dependent signalling pathways. These transitions arise in this model purely from the physical interactions and the curvature-force coupling.

To test these predictions, we examined phagocyte--target interactions across multiple experimental systems, including macrophage engulfment of Giant Unilamellar Vesicles and lymphoma cells, together with epithelial clearance of apoptotic targets \emph{in vivo} using synthetic targets of defined stiffness. Across all systems, the observed behaviours closely matched the predicted stiffness-dependent transitions between biting, pushing, and engulfment. In particular, we validate the predicted intermediate ``pushing'' regime between trogocytosis and complete phagocytosis and show that it arises from curvature-dependent recruitment of actin at the cell membrane. These results identify a general physical mechanism by which target mechanics regulates phagocyte--target interactions across diverse cell and target types. Together, our findings establish target stiffness as a key regulator of cellular clearance during immune surveillance and tissue homeostasis.\\

\section*{Theoretical Model}\label{sec:model}
\begin{figure}
    \centering
    \includegraphics[width=\linewidth]{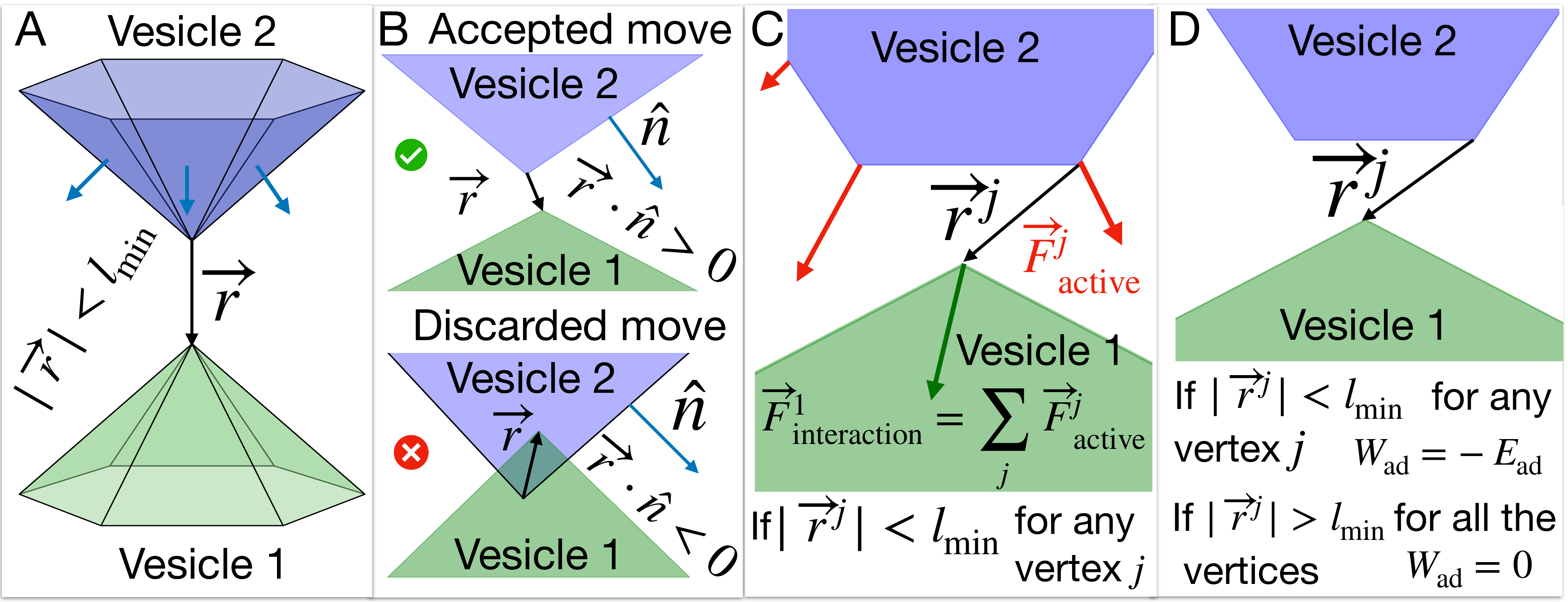}
    \caption{Interaction between two vesicles in the model: A) Two surfaces of two vesicles are shown in green and blue colors respectively. We check the distance between the vertices that belong to the two different vesicles and determine it they are interacting if the distance between them is less than the length unit of our simulation, i.e., $l_{\rm min}$. B) If the distance                   $|\protect\overrightarrow{r}|$ is less than $l_{\rm min}$, we find the dot product between the vector $\protect\overrightarrow{r}$ and the normals of all the triangles common to the vertex from the other vesicle $\protect\hat{n}$. If the dot product is negative, the MC move is discarded as the vesicles are overlapping; otherwise, it is accepted. C) The interaction force on the vertex of interest is the vector sum of the active forces applied by the vertices on the other vesicle within the interaction range. D) If a vertex from vesicle 1 is at a distance less than the interaction range $l_{\rm min}$ from any vertex of another vesicle, then the adhesive energy between them is $-E_{\rm ad}$ for both of these vertices. Otherwise, it is zero.}
    \label{fig:interaction_schematic}
\end{figure}

Our theoretical model is based on the Monte-Carlo (MC) calculation of the dynamics of a closed three-dimensional triangulated self-avoiding vesicle with a spherical topology (Fig.\ref{fig:interaction_schematic})~\cite{Fonari2019,sadhu2021modelling,sadhu2023minimal} (See SI sections S1-S3 and Method and material for details). Within this model we denote the bare membrane nodes in blue, and the nodes containing the CMC in red. The active protrusive forces, representing the result of actin polymerization, are applied in the direction of the outwards local normal, at the locations of the CMC. In the present model we do not describe any contractile forces that can result from the activity of myosin-II motors. While myosin-II contractility affects the efficiency of phagocytosis and mostly gets activated during the late stages of the process, it is not an essential component \cite{sosale_2015_cell,Vorselen2021_elife, Jaumouill2019_Nat_cell_bio}. These observations allow us to focus here on the regulation of the crucial recruitment of the actin cytoskeleton and its resultant protrusive forces, which dominate the phagocytosis process. Future extensions of our model will explore the role of contractile forces.

Our model is purely a membrane model, and it does not include any information about the details of the actin network inside the cell or the internal structure of the cell, such as organelles (nucleus) or bulk elastic deformations and their associated energy cost.

Here we developed our previous model to allow for the interaction between two dynamic vesicles (Fig.\ref{fig:interaction_schematic}). As a first step, we find the nodes that are adjacent between the two vesicles (Fig.\ref{fig:interaction_schematic}A). Such proximal nodes are restricted in their MC moves, as we do not allow the two vesicles to pass through each other. We need to check any such pair of vertices from different vesicles (Fig.\ref{fig:interaction_schematic}B, Fig.S1).

Next, we consider the active forces that a vesicle exerts on its neighboring vesicle. Any vertex in vesicle $1$ feels the active force due to the other vesicle's active CMC sites, that are within the interaction range (Fig.\ref{fig:interaction_schematic}C). The effect of this active force is included by adding the energy term given by,
\begin{equation}
    W_{\rm int}=F\sum_i \hat{n}_{i}\cdot \overrightarrow{dr}
\end{equation}
where $i$ runs through all the vertices belonging to vesicle $2$ within the distance of $l_{\rm min}$ from the vertex of interest in vesicle $1$, the force vectors have amplitude $F$ and are directed at the outwards normal $\hat{n}_{i}$ at the sites $i$, while $\overrightarrow{dr}$ is the MC move of the vertex in vesicle $1$. Note that in our model we do not explicitly maintain force balance. When adhered to external surfaces, they serve as momentum sinks, while for free vesicles we work in the center-of-mass frame which allows us to calculate relative shape changes, such that globally the force is effectively balanced.

Finally, we introduce an adhesion energy between proximal vertices on the two neighboring vesicles. Each vertex $j$ that is within the adhesion range to the other vesicle (Fig.\ref{fig:interaction_schematic}D) has an adhesion energy that is given by,
\begin{equation}
    W_{\rm ad}^j=-E_{\rm ad}
    \label{eq:adhesion_energy_term}
\end{equation}
Initially, we evolve it for a few Monte-Carlo steps from a pentagonal-dipyramid, until it becomes nearly spherical. The active force (when implemented) is set to $F=2~k_BT/l_{\rm min}$ and the protein-protein interaction energy is set to $w=1~k_BT$ throughout the paper.

\begin{figure}
    \centering
    \includegraphics[width=\linewidth]{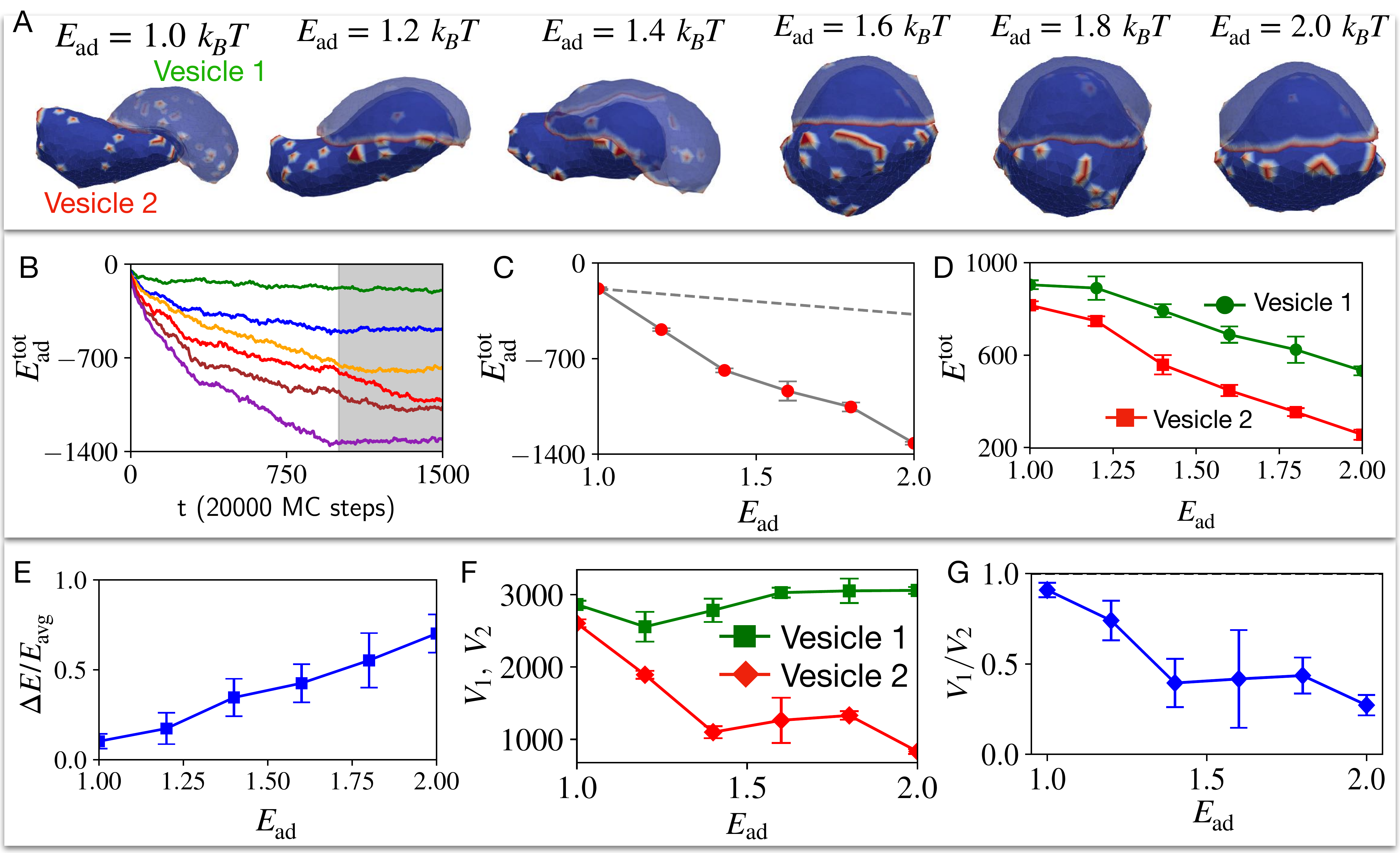}
    \caption{Two identical vesicles adhere to each other for different strengths of adhesion energy parameter $E_{\rm ad}$. A) Final configuration snapshots of the pair of passive vesicles (at time step $t=1500$), for the adhesion energy parameter $E_{\rm ad}=1,~1.2,~1.4,~1.6,~1.8,~2.0$ in units of $k_BT$. Blue denotes the bare membrane nodes, while red at the passive CMC nodes. B) The average adhesive energy $E^{\rm tot}_{\rm ad}$ per vesicle is shown as a function of timestep for $E_{\rm ad}=1,~1.2,~1.4,~1.6,~1.8,~2.0$ in units of $k_BT$ with green, blue, orange, red, brown and purple solid lines respectively. C) The final total adhesive energy $E^{\rm tot}_{\rm ad}$ (averaged over the grey shaded time window shown in (B)) is shown for six different values of $E_{\rm ad}$. The grey dashed line represents the total adhesive area in the case of $E_{\rm ad}=1~k_BT$ multiplied by the $E{\rm ad}$. It shows that the total adhesive energy increases due to the increase in adhered area, faster than the increase in $E_{\rm ad}$ (dashed line). In D) we show the total energies the pair of vesicles, averaged over the grey shaded region for six different values of $E_{\rm ad}$. E) The relative difference between the vesicles increases with $E_{\rm ad}$. F) Average volumes of the vesicles as function of $E_{\rm ad}$, and G) the corresponding volume ratio.  We used $722$ vertices for each vesicle of bending rigidity $\kappa=20~k_BT$, out of which $50$ vertices represent the curved-protein complexes with intrinsic curvature $c_0=1~l_{\rm min}^{-1}$, i.e. the CMC percentage is $\rho=6.93~\%$. Here, volume is not conserved.}
    \label{fig:symmetry_break}
\end{figure}

The area of the vesicles in our model is roughly conserved, with the bond lengths limited in their allowed range of fluctuation ($l_{\rm min}<l<1.7 ~l_{\rm min})$ \cite{Fonari2019}. The typical changes in area in our simulations (without applied osmotic pressure) are of order $5~\%$ (see Fig.S7). When the osmotic pressure is applied inside the vesicle, the bond lengths are pushed towards their maximal allowed value (and therefore the average area increases), and the fluctuations in area are greatly suppressed. Finally, we note that the MC method does not include dissipative processes such as membrane-membrane and viscous friction, and therefore do not give information about the physical time-scale of the simulated process.

\section*{Results}

\subsection*{Spontaneous symmetry breaking of adhering vesicles containing CMC}

We first analyzed the adhesion between two identical vesicles, in the absence of active forces ($F=0$). For calibration purposes, we tested our model for adhesion of bare membrane vesicles (no CMC) and under the conditions of volume conservation (Figs.S2-S4, See SI section S6), for which analytic solutions are available \cite{pavlivc2009encapsulation}. In this limit, we could extract the contact angles between the two vesicles, and compare them to an analytic result \cite{tordeux2002analytical,frank2008prevention,marevs2012determination}. The agreement between the simulations and the analytic calculation serves to validate our numerical procedure. Adding CMC induces larger spreading and some breaking of the rotational symmetry, though the two adhering vesicles remain symmetric with respect to each other (Fig.S4). In the presence of CMC the contact surface between the vesicles is not anymore composed of flat or spherical surfaces, as was found for simple adhered vesicles \cite{urbanija2008attachment,pavlivc2009encapsulation}.

When we remove the volume conservation constraint (See SI section S7), we find that the presence of CMC can drive a spontaneous symmetry breaking above a critical value of the adhesion energy (Fig.\ref{fig:symmetry_break}A). At low adhesion strength the two vesicles are still symmetric (Fig.\ref{fig:symmetry_break}A, $E_{\rm ad}=1 ~k_BT$), but above a critical adhesion strength one of the vesicles spontaneously forms a cup-like shape (top vesicle in Fig.\ref{fig:symmetry_break}A, vesicle 2), which partially encapsulates the other vesicle (bottom vesicle in Fig.\ref{fig:symmetry_break}A, vesicle 1). The CMC in the top vesicle condense along the sharp rim of the cup shape, similar to the organization of such a vesicle when engulfing a rigid sphere \cite{sadhu2023theoretical} (Fig.\ref{fig:passive}A). The bottom vesicle remains largely spherical, with the CMC randomly spread as small isolated clusters. This transition is driven by a lowering of the total energy of the two vesicles. The bending energy increases during the symmetry-breaking transition (Fig.S5), as the cup-shaped vesicle $2$ is highly curved. However, this increase is offset by the adhesion energy between the vesicles which decreases the total energy (Fig.\ref{fig:symmetry_break}B-E), and to a much smaller amount by the CMC-CMC binding energy (Fig.S5). The huge changes to the volume of the cup-shaped vesicle $2$ during this spontaneous shape transition are quantified in Fig.\ref{fig:symmetry_break}F,G.

Shapes that are similar to the  vesicle pair of Fig.\ref{fig:symmetry_break}(A) were obtained for adhering soft tissues (modeled as vesicles), where the symmetry breaking was induced by a large difference in the active surface tension between the two vesicles \cite{torres2022interacting}. In our system the symmetry breaking is spontaneous, and the two vesicles have identical properties.

\subsection*{Rigidity-dependent biting, pushing and engulfing}

Next we explore the process of engulfment that mimics the phagocytosis of a non-rigid object by a cell. The engulfed (target) object is described by a small vesicle ($N^T=847$ vertices, forming a spherical object with radius of $10~l_{\rm min}$), and the cell-like (bigger) vesicle has $3127$ vertices. We set the CMC concentration $\rho =4.8~\%$ on the cell-like vesicle throughout this work. We start by validating our computation, demonstrating that for a very rigid target vesicle (bending modulus $\kappa=2000~k_BT$) the engulfment proceeds in the same manner as we previously computed for a perfectly rigid sphere (Fig.S6, and see SI section S8) \cite{sadhu2023theoretical}. Note that we are using the bending modulus as a model parameter to easily control the rigidity of the target vesicle, with a higher bending modulus diminishing shape deformations that increase the local mean curvature.

\begin{figure}
    \centering
    \includegraphics[width=\linewidth]{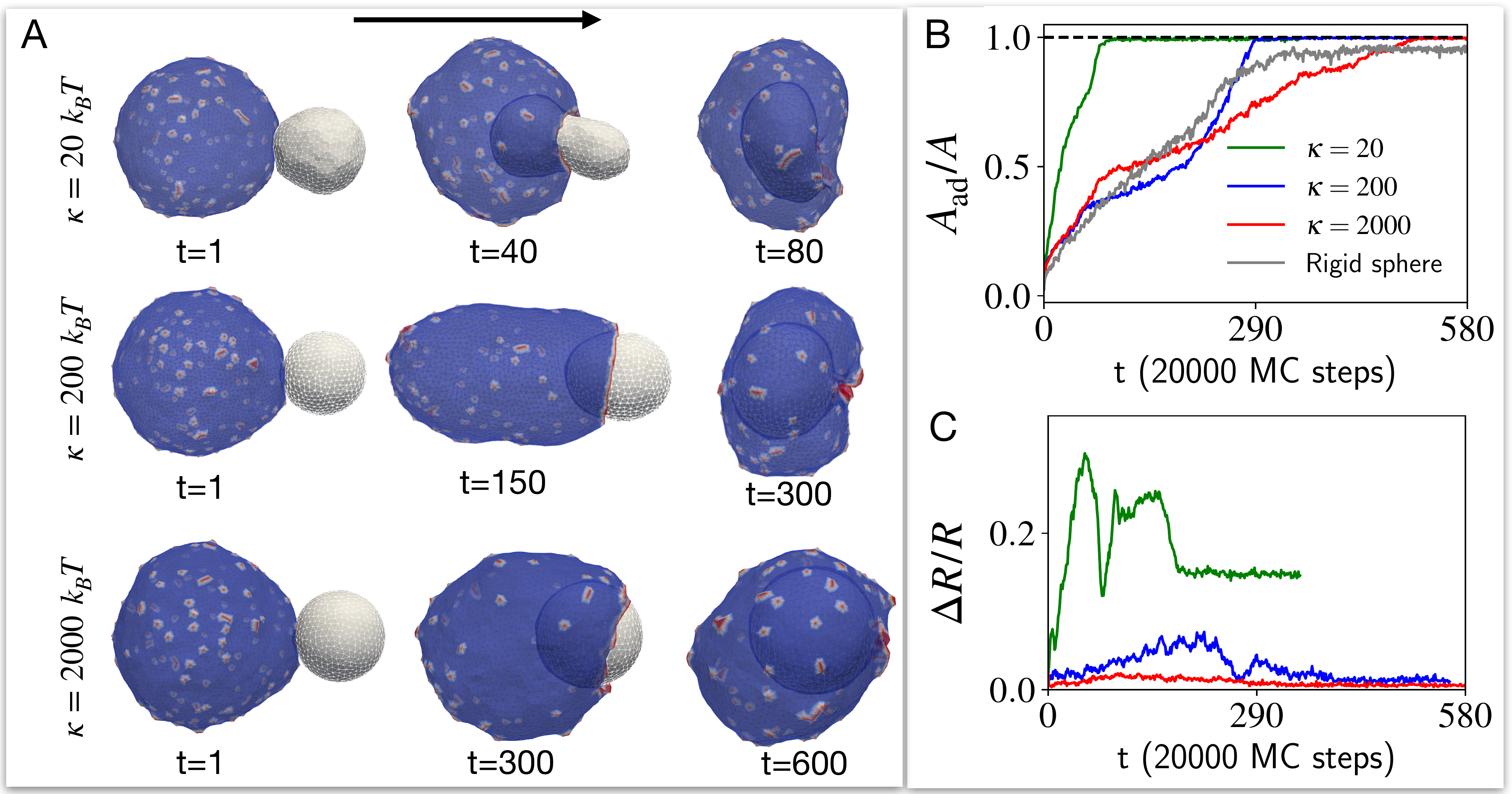}
    \caption{The engulfment of target vesicles of different rigidities by a cell-like vesicle with passive CMC. A) The snapshots of the shapes of the interacting vesicles with time for different bending rigidity $\kappa$ values of the target vesicle. We set the bending rigidity of the cell-like vesicle at $20~k_BT$. B) The time evolution of adhered area fraction of the target vesicle for different $\kappa$ values (see SI section S4 for details on this calculation). As $\kappa$ increases, the outcome is approaching the completely rigid $\kappa=\infty$ limit that is shown in grey. C) Time evolution of the deviation of the target vesicle from a spherical shape, for different values of $\kappa$ (color code as in (B)), measured by the relative standard deviation for the position vector of all the vertices with respect to the center of mass of the vesicle (Eq.S11). As $\kappa$ increases, the target remains spherical all the time, and it takes longer time to engulf as it requires the cell-like vesicle to extend its adhesion cup over a bigger cross-section.}
    \label{fig:passive}
\end{figure}

\begin{figure*} 
    \centering
    \includegraphics[width=1\linewidth]{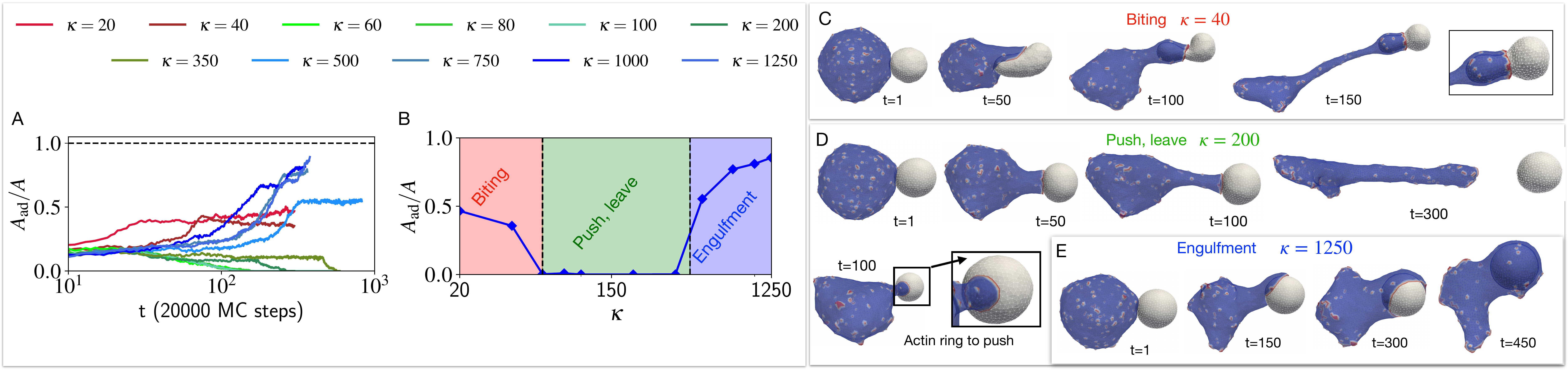}
\caption{Effect of the bending rigidity of the target vesicle on the process of phagocytosis. The time evolution of the adhered area fraction of the target cell is shown in (A), for different bending modulus values. B) The final adhered area fraction of the target vesicle is shown as a function of the bending rigidity $\kappa$. We indicated three phases of biting, push-leave, and engulfment with red green and blue as $\kappa$ increases. The time evolution of the shapes and the snapshots are shown for three example of biting, push-leave, and engulfment in C), D) and E) panels, where the bending rigidity $\kappa$ for the target vesicle is set to $40~k_BT,~200~k_BT,~$ and $1250~k_BT$ respectively. We set the bending rigidity of the cell-like vesicle to $20~k_BT$.}
    \label{fig:kappa_transition_part1}
\end{figure*}

We start exploring how a cell-like vesicle that contains passive CMCs ($F=0$, Fig.\ref{fig:passive}A), engulfs a target vesicle of different bending modulus. We find that the cell-like vesicle with passive CMC is able to fully engulf the target vesicle, with the engulfment proceeding faster for the softest target vesicle (See Movie S1). This faster engulfment is facilitated by the large deformation (See SI section S9 for details of deformation measurement) of the target vesicle (as shown in Fig.\ref{fig:passive}B,C), which enables the cell-like vesicle to extend an adhesion cup over a smaller cross-sectional area.

This behaviour is drastically changed when the CMCs induce active protrusive forces. We explore in Fig.\ref{fig:kappa_transition_part1} the engulfment dynamics as function of the bending modulus $\kappa$ of the target vesicle, and find three main dynamical phases. We start with the high $\kappa$ regime (blue traces in Fig.\ref{fig:kappa_transition_part1}A, and blue region on the phase diagram Fig.\ref{fig:kappa_transition_part1}B), where the cell-like vesicle completely engulfs the rigid target vesicle (similar to the engulfment of a rigid object, Fig.S6 \cite{sadhu2023theoretical}), as shown in the snapshots of Fig.\ref{fig:kappa_transition_part1}E. For a softer target vesicle (green traces in Fig.\ref{fig:kappa_transition_part1}A, and green region on the phase diagram Fig.\ref{fig:kappa_transition_part1}B), we find that the target vesicle ends up being pushed away (Fig.\ref{fig:kappa_transition_part1}D), and the contact area stalls in a ``suction-cup"-like shape, and later retracts (Fig.\ref{fig:kappa_transition_part1}A) until the two vesicles detach (zero final adhered area fraction, Fig.\ref{fig:kappa_transition_part1}B) due to the high bending energy of the elongated ``finger" attached to the target vesicle (Fig.S8, See SI section S10). At even lower values of $\kappa$ (red traces in Fig.\ref{fig:kappa_transition_part1}A, and red region on the phase diagram Fig.\ref{fig:kappa_transition_part1}B) we find that the engulfed area stalls at a small value ($A_{ad}/A<0.5$, Fig.\ref{fig:kappa_transition_part1}A,B), corresponding to a ``biting"-type dynamics (Fig.\ref{fig:kappa_transition_part1}C). Note that we do not allow the vesicles to undergo fission, even when greatly deformed.

\begin{figure*}
    \centering
\includegraphics[width=1\linewidth]{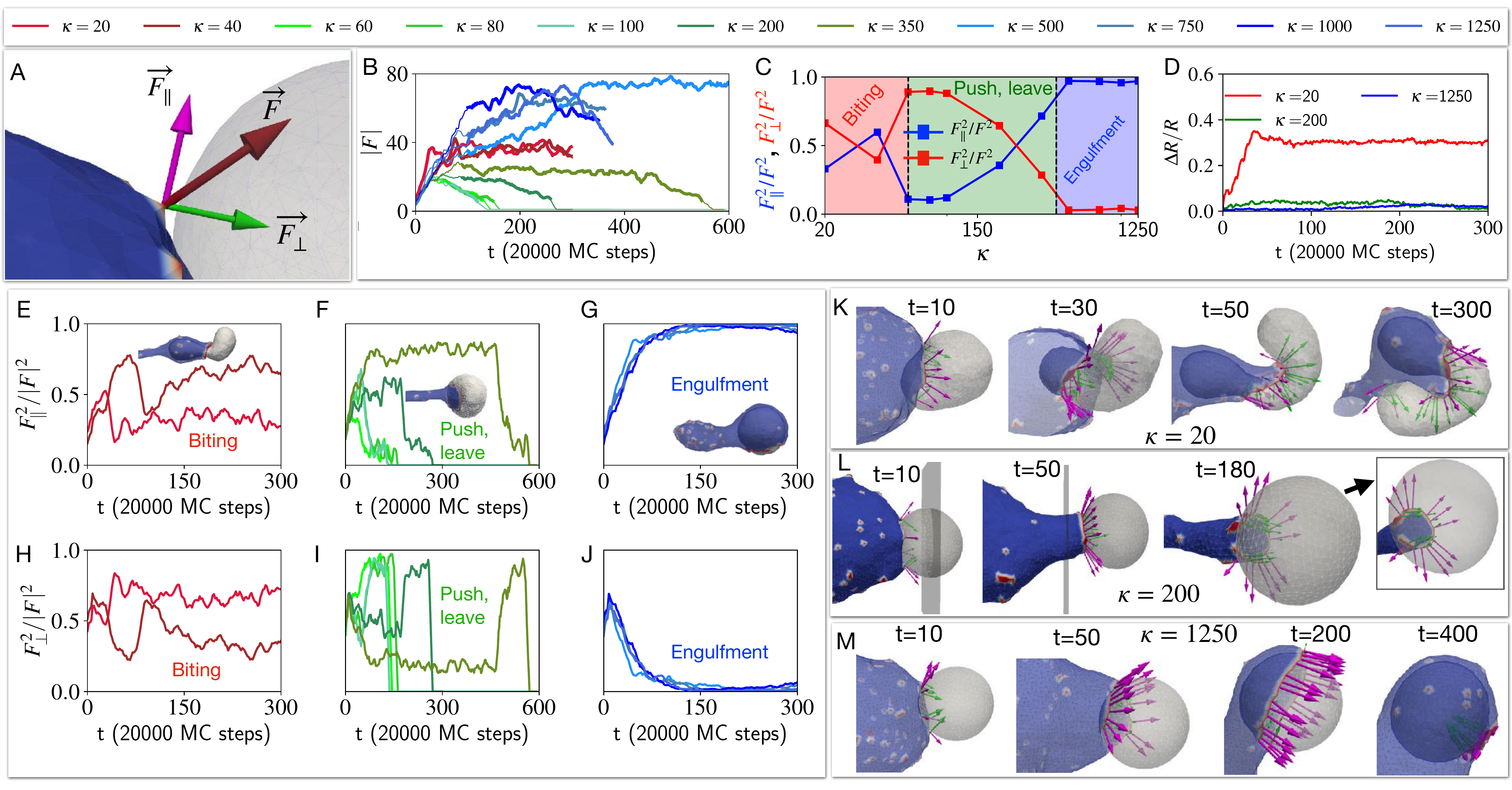}
    \caption{Coupling between the target vesicle deformation and alignment of the active forces during the engulfment process. A) The active force due to an active CMC node of the cell-like vesicle is applied to a neighboring node of the target vesicle $\protect\overrightarrow{F}$. This force is decomposed into two parts, $\protect\overrightarrow{F}_{\parallel}$ is tangential to the surface of the target vesicle and $\protect\overrightarrow{F}_{\perp}$ is the pushing normal force (shown in magenta and green colour, respectively). B) The time evolution of the total magnitude of the force applied to the target vesicle by the cell-like vesicle is shown, for different bending rigidities of the target vesicle (as in Fig.\ref{fig:kappa_transition_part1}). C) Time averaged the tangential and normal force fractions as function of the bending rigidity of the target vesicle $\kappa$ (over the times denoted by bold lines in (B)). D) The deviation of the target vesicle from a sphere is shown for three different bending rigidities $\kappa$ of the target vesicle (Eq.S11). E)-G) show the fraction of tangential force on the target in the three dynamical regimes of biting, push-leave, and engulfment. H)-j) Similarly, the fraction of normal pushing force on the target in the three dynamical phases. K)-M) Snapshots showing the force decomposed into tangential and normal components together with the deformations of the target vesicle. We show how the early deformation caused by the normal components affect the later alignment of the CMC at the leading edge of the adhesion patch. K) A very soft target vesicle can initially adhere and bend into the cell-like vesicle, during the early stages. However, the CMC then impinge against the remaining target vesicle, and end up pushing and twisting it with a significant normal component. L) In the pushing regime the deformation of the target vesicle is sufficient to prevent the CMC from aligning tangentially, and a significant normal component maintains the pushing dynamics. M) For the rigid target vesicle the CMC cluster aligns tangentially and drives efficient engulfment. The bending rigidity of the bigger cell is set to $20~k_BT$.}
    \label{fig:kappa_transition_part2}
\end{figure*}

The origin of these dynamical phases can be understood when we investigate how the membrane shape and the orientation of the active forces are coupled. In Fig.\ref{fig:kappa_transition_part2}(A) we define the components of the active force that is exerted by a CMC of the cell-like vesicle when its in contact with the nodes of the target vesicle. This force is applied towards the outwards normal of the CMC, and has both normal ($F_{\perp}$) and tangential ($F_{\parallel}$) components with respect to the target vesicle  surface. 

During the engulfment phase of rigid target vesicles we find that the CMC form large leading-edge clusters (large value of the total transmitted force, blue traces in Fig.\ref{fig:kappa_transition_part2}B) which mostly exert tangential forces (Fig.\ref{fig:kappa_transition_part2}G,J). This is shown in the snapshots of Fig.\ref{fig:kappa_transition_part2}M. The average fraction of the two force components (beyond the transient initial cup formation stage, denoted by the bold lines in Fig.\ref{fig:kappa_transition_part2}B
), is shown in Fig.\ref{fig:kappa_transition_part2}C, and clearly demonstrate the relation between these force components and the resulting three dynamical phases.

In the pushing phase of slightly softer target vesicles we find that the leading-edge cluster gets arrested at a smaller size (green traces in Fig.\ref{fig:kappa_transition_part2}B), eventually disappearing when the two vesicles disengage. The dynamics of the force components (Fig.\ref{fig:kappa_transition_part2}F,I) show that the normal component remains relatively high, preventing the efficient spreading of the engulfing membrane over the target vesicle. In Fig.\ref{fig:kappa_transition_part2}L the snapshots show the origin of this behavior: due to the deformation of the target vesicle, the leading-edge of the engulfing membrane is not tangentially oriented, further pushing into the target vesicle and maintaining its deformation. This feedback between shape and force orientation arrests the spreading, as the target vesicle is pushed and deformed.

In the regime of softest target vesicle (red traces in Fig.\ref{fig:kappa_transition_part2}B) the target vesicle gets strongly deformed by the adhesion to and active forces exerted by the engulfing membrane (red line in Fig.\ref{fig:kappa_transition_part2}D). This large deformation prevents the CMC and the active force from aligning tangentially, with both components having similar magnitude (Fig.\ref{fig:kappa_transition_part2}E,H). Due to the large deformations of the soft target vesicle (Fig.\ref{fig:kappa_transition_part2}K) the leading-edge ends up mostly pushing it after a small portion is engulfed, leading to a ``biting"-like behavior. This is very different from the smooth engulfment observed when the active forces are absent (Fig.\ref{fig:passive}).

In Fig.S9 we show the same set of dynamic phases when we vary the osmotic pressure (see SI section S12 for details) inside the target vesicle (while keeping $\kappa=20~k_BT$). We find that as the internal pressure $p$ inside the target vesicle increases, the membrane of this vesicle gets stretched out and tense, inhibiting any shape changes and effectively stiffening the vesicle. Not surprisingly, the observed phases, as function of the pressure $p$ (Fig.S9), perfectly match the phases observed as function of bending rigidity in Fig.\ref{fig:kappa_transition_part1}. 

To summarize, in our model the recruitment and organization of the actin-derived forces at the leading edge of the cell are curvature-sensitive. This curvature-activity coupling makes the forces' alignment dependent on the target deformation, and gives rise to three dynamical phases as function of the target's rigidity. These theoretical predictions, especially the emergence of an intermediate ``pushing" phase, will next be compared to experimental observations.

\subsection*{Comparison to experiments}

We start by validating our model for the localization of actin polymerization at the leading edge of the phagocytic cup, and how it transmits forces to the engulfed object. We compare to experiments utilizing elastic beads \cite{vorselen2020microparticle,settle2024beta2}, which enable the extraction of the forces exerted by the cell during the phagocytosis process (Fig.S12). Since the beads are composed of a cross-linked gel that has bulk elasticity, we roughly emulate it by adding an effective spring network to our membrane (details in SI section S13). This is a rough approximation, compared to detailed theoretical descriptions of cross-linked gels engulfed by membranes \cite{sorokina2024computer,debnath2025wrapping}. Nevertheless, the comparison shows very good overall similarity in both the deformations of the beads, and the associated localization of the normal forces exerted by the cell’s leading edge. This good agreement validates that protrusive forces in the model are exerted realistically both spatially and temporally, supporting our association of the actin force with the highly curved CMC cluster in the simulations. A band of normal force exerted by this leading edge on the engulfed sphere results in local squeezing of the sphere, in both experiments and simulations. We also observe that the actin at the leading edge is non-uniform and fragmented in both the experiments and our simulations, suggesting that a simple coupling between curvature and actin nucleation may be involved in a complex and non-uniform organization of this moving front \cite{herron2022actin,sopelniak2025phagocytic} due to its sensitive dependence on the target deformations.

Following the validation of the forces exerted in our model, we now compare to experimental data showing the dependence of the phagocytosis dynamics on target stiffness. Our model's prediction of a ``pushing" phase for intermediate target stiffness offers an explanation to the puzzling observations of cells either engulfing or pushing away apoptotic cells during embryogenesis \cite{hoijman2021cooperative}. To directly test this prediction of our model that the mechanical state of apoptotic targets determines whether engulfment or pushing behaviours take place, we used the early zebrafish embryo as an in vivo model system. Epithelial cells in the early embryo have previously been shown to generate two distinct types of protrusions: phagocytic cups that mediate apoptotic cell engulfment and “epithelial arms” that exert pushing forces and promote apoptotic cell displacement \cite{hoijman2021cooperative}. We therefore aimed to assess if target stiffness may represent a key biophysical cue governing the choice between these two modes of phagocyte-target interaction. For this purpose, we employed synthetic apoptotic targets with tuneable mechanical properties that recapitulate the lipid composition and phosphatidylserine (PS) exposure of apoptotic cells (termed lipobeads in the following). Using quantitative live in vivo imaging, we first confirmed that soft lipobads (1.3 kPa) versus stiff lipobeads (80 kPa) were efficiently recognized and internalized by epithelial cells (Fig. S11A,D). High-resolution imaging further captured individual uptake events, demonstrating bona fide phagocytic engulfment of lipobeads by the epithelial tissue (Fig. S11B,C).

Notably, whereas stiff lipobeads (80 kPa) were exclusively engulfed by epithelial cells (Fig. S11A), soft lipobeads (1.3 kPa) exhibited two distinct outcomes: they were either actively displaced by epithelial cells through pushing interactions (Fig. 6A) or subsequently engulfed (Fig. 6B). Consistent with these observations, quantitative tracking revealed a pronounced increase in target mobility for soft compared with stiff lipobeads (Fig. 6C,D; Movie S12). Apoptotic cells generated through overexpression of the pro-apoptotic factor Bax displayed the highest degree of mobility, in agreement with a softening of apoptotic cells \cite{Pelling2009,VanderMeeren2020}. Together, these results provide direct in vivo evidence that apoptotic target stiffness is a major determinant of epithelial protrusion dynamics and thus clearance behaviour. Soft lipobeads and apoptotic cells were frequently observed to be displaced by epithelial cells and subsequently detach from the pushing cell (Fig. 6A)\cite{hoijman2021cooperative}, as predicted by our simulations (Fig. 4D). 

Engulfment of soft lipobeads was observed when the target became mechanically constrained by neighbouring epithelial cells (Fig. 6B), suggesting that local tissue geometry can modulate the outcome of cell–target interactions. To investigate the role of spatial constraints, we incorporated target confinement into our model by fixing a small region of the target vesicle in space (black patch at the pole opposite to where the cell is at t=1, Fig. 6E,F). This constraint has a substantial effect on phagocyte-target interaction, resulting in engulfment (Fig.6G), as it allows the protrusive forces at the leading edge of the cell-like vesicle to align tangentially to the target (Fig.6H,I), which is needed to drive efficient engulfment. In contrast, unconstrained soft targets are displaced before stable wrapping can be established, resulting in pushing-mediated target transport rather than uptake.

In Fig.\ref{fig:three_behavior} we compare the theoretical predictions of three dynamical phases as function of the membrane tension of the target vesicle, to experimental observations. Fig.\ref{fig:three_behavior}(A-C) gives examples of snapshots of the interactions between macrophage cells (labeled in green) and  giant unilamellar vesicles that contain antibodies that trigger macrophage adhesion (GUVs, labeled in pink) \cite{cornell2025target}. In the experiments the membrane tension was varied using  different sucrose solutions to fill the GUVs, and as function of decreasing membrane tension the observed behavior changes from (mostly) engulfment at high tension (Fig.\ref{fig:three_behavior}A), to a mixture of pushing and some biting at intermediate tensions (Fig.\ref{fig:three_behavior}B), finally exhibiting biting (trogocytosis) activity for the lowest GUV membrane tensions (Fig.\ref{fig:three_behavior}C, See Movies S2, S3, S4).

Our model and observations suggest that the engagement of target populations that display heterogeneous stiffnesses, such as cancer cells, could lead to a mix of pushing, partial and complete phagocytosis. To assess this possibility with physiological targets, we performed time-lapse microscopy of macrophages (green) interacting with lymphoma cells opsonized with therapeutic antibodies  (magenta) (Fig.\ref{fig:three_behavior}D-F).
From top to bottom we see the same three typical behaviors: engulfment (Fig.\ref{fig:three_behavior}D), pushing (Fig.\ref{fig:three_behavior}E) and biting (trogocytosis, Fig.\ref{fig:three_behavior}F) (See Movies S5, S6, S7 respectively). While we do not know exactly the stiffness of each of these cancer cells, finding this range of behaviors and comparing to our model suggests that cancer cells exhibit a wide range of stiffness \cite{rianna2020direct}. This trait of cancer cells was indeed measured using Atomic Force Microscopy (AFM) \cite{lekka2016discrimination,yang2025single}.

In Fig.\ref{fig:three_behavior}(G-I) we demonstrate how the three phases observed in the experiments are predicted by our theoretical simulations as a function of the pressure inside the engulfed target vesicle (See Movies S8, S9, S10, respectively). Higher internal pressure acts to suppress dynamic changes and fluctuations in the target vesicle's area, equivalent to higher membrane tension (Fig.S10). Note that we do not allow our vesicles to undergo topological changes such as fission, so the biting behaviour in the simulations is arrested.

\section*{Discussion}

Our combined experimental and theoretical study identifies target mechanics as a previously underappreciated regulator of phagocyte behaviour, with relevance for the clearance of opsonized targets, pathogens and apoptotic cells. While target recognition and uptake has primarily been addressed from molecular receptor-ligand signalling, our findings strengthen increasing evidence that the mechanical target state provides critical information that influences the mode of a phagocyte’s response. 

We have explored here the dynamics of cellular adhesion and engulfment of soft objects, whereby the mechanism of the self-organization of the protrusive forces of actin polymerization is through curved membrane protein complexes (CMC). Curved membrane proteins that recruit actin polymerization were shown to drive cellular protrusions during cell adhesion and migration \cite{begemann2019mechanochemical,wu2025wave,sadhu2023minimal}. This mechanism was previously shown to explain how cells phagocytose rigid objects \cite{sadhu2023theoretical}, and here we explored this mechanism when the engulfed object is flexible. 

Our theoretical model of phagocytosis goes beyond passive models of engulfment driven purely by adhesion \cite{tollis2010zipper}, by describing the role of active cytoskeletal forces. A previous work that included active cytoskeletal forces \cite{zhang2024signaling} also found engulfment rate that diminishes when the target is softer, due to the target deformation affecting the cell's ability to spread by adhesion and random active force. In our model we provide an explicit mechanism that explains how the target deformation affects the organization of the cell's cytoskeleton, by coupling the recruitment of actin polymerization to the shape of the cell and the target, through curvature-sensitive membrane protein complexes.

\begin{figure}
    \centering
    \includegraphics[width=\linewidth]{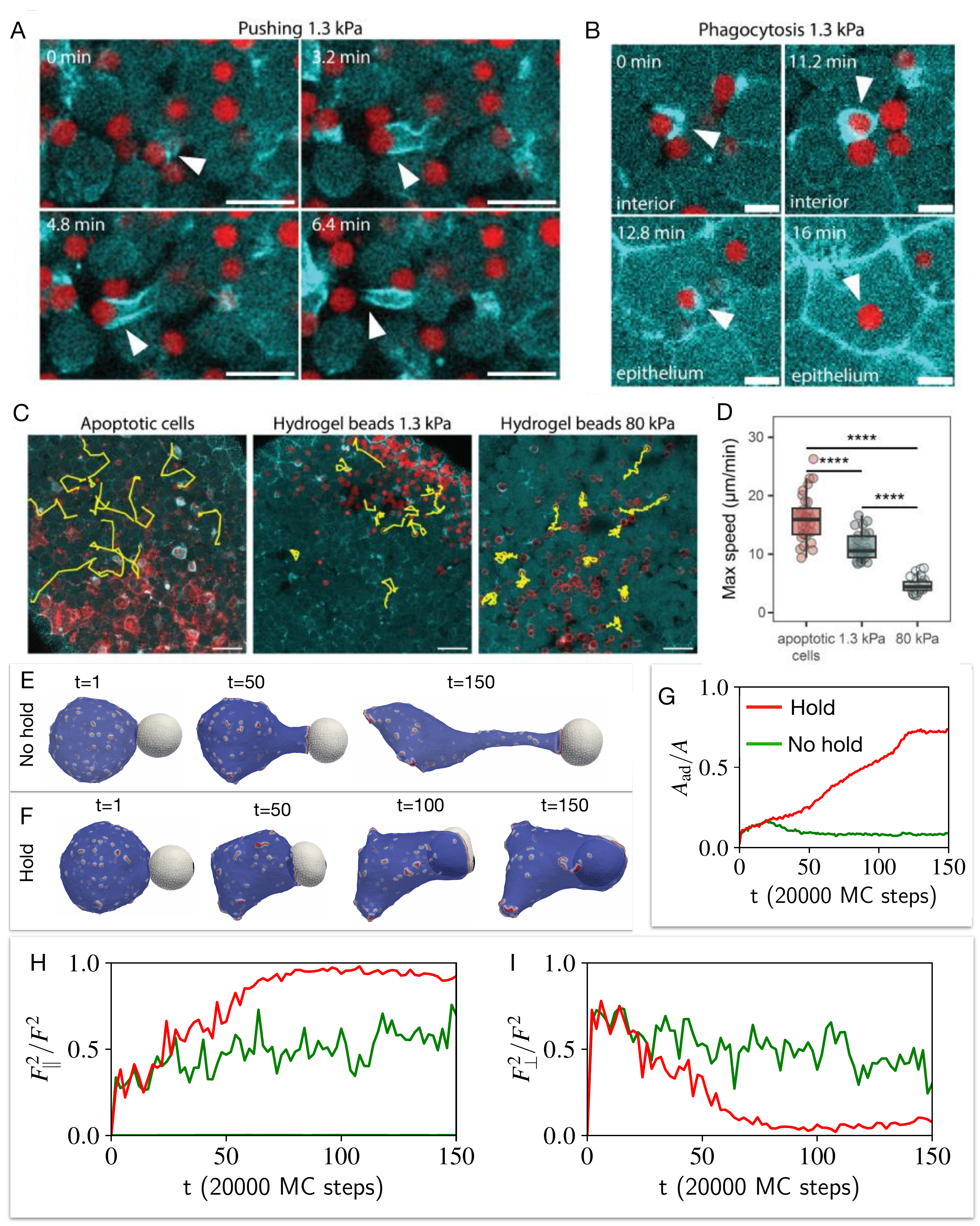}
    \caption{Epithelial phagocytosis and mobility of apoptotic cells and synthetic targets with variable target stiffness in vivo. A. Representative images of a pushing event of 1.3 kPa hydrogel beads over the indicated time period in vivo. Arrowheads point at the formation of an epithelial pushing protrusion termed ‘epithelial arm’, which leads to the movement of the synthetic apoptotic target in vivo.  B. Representative images of an individual phagocytosis event (arrowhead) of 1.3 kPa hydrogel beads in vivo. C. Representative x/y-trajectories (yellow lines) obtained from 3D tracking of apoptotic cells (Bax+ cells, red, left), 1.3 kPa hydrogel beads (red, middle); and 80 kPa hydrogel beads (red, right) in embryos expressing Lifeact-GFP (cyan). D. Analysis of maximum target speed for apoptotic cells (red, n = 32 tracks from 3 embryos), 1.3 kPa hydrogel beads (dark cyan, n = 29 tracks from 3 embryos) and 80 kPa hydrogel beads (light cyan, n = 30 tracks from 3 embryos). Data points represent individual tracks. Pairwise comparisons using Wilcoxon rank sum test $p<0.0001$ between all conditions. All embryos were obtained from the Tg(actb1:Lifeact-GFP) line. Size of hydrogel beads: 8.8$\mu$m (1.3 kPa), 8.2$\mu$m (80 kPa).  E. Shows the snapshots of cell-like vesicle pushes the target away when the target is not held by any external means. F. Shows the snapshots of how the cell-like vesicle starts to engulf the target while it is held by some external means (See SI section S11). H. The fraction of force applied by the cell-like vesicle parallel to the target's surface. I. The fraction of force applied by the cell-like vesicle normal to the target's surface. Scale bars: 20$\mu$m (A), 10$\mu$m(B), 40$\mu$m (C).}
    \label{fig:verena}
\end{figure}
Our model predicts that as function of the stiffness or membrane tension of the engulfed vesicle, there are distinct dynamical phases: While a stiff object is engulfed, a softer object will be spontaneously pushed away until it detaches. The softest objects get partially engulfed, which we expect in reality to result in a piece getting ``bitten" off (trogocytosis). The theoretical model explains the origin of these dynamical phases as arising due to the feedback between the deformations of the engulfed object, and the orientation of the active forces exerted by the cell's leading edge, due to the curvature-activity coupling. Comparing to experiments of in-vivo epithelial phagocytosis of apoptotic cells and artificial beads, as well as in-vitro immune cells engulfing artificial vesicles or cancer cells, we validate the predicted relation between engulfment dynamics and the mechanical deformability of the target. Crucially, the experiments validate the theoretically calculated ``pushing" phase, which was predicted and observed to emerge at intermediate target stiffness. 

Here we show that epithelial cells selectively push soft versus stiff targets prior to engulfment. In the context of apoptotic target clearance by epithelial cells, this increased pushing of apoptotic targets thereby facilitates a cooperative clearance that reduces the overall time it takes for epithelial cells to remove apoptotic cells from the tissue \cite{hoijman2021cooperative}. Furthermore, our data suggest that local tissue geometry can modulate these mechanically driven outcomes, as confinement by neighbouring cells promotes engulfment of otherwise displaceable soft targets. The observed dependence of phagocyte dynamics on target stiffness is particularly intriguing considering mechanical changes that accompany apoptosis. During apoptosis progression, mechanical structures within the cell are modified, leading to changes in cortical tension, cytoskeletal organization and cellular stiffness \cite{Pelling2009,VanderMeeren2020}. Our results raise the possibility that such mechanical remodelling is not merely a consequence of cell death but can actively influence how dying cells are handled by phagocytic cells.

Our theoretical results offer an explanation of the lower rates of successful engulfment observed for soft beads \cite{beningo2002fc,Vorselen2020_nat_comm,settle2024beta2}, in agreement with our simulation results (Fig.\ref{fig:kappa_transition_part1}). Similarly, red-blood cells were observed to be phagocytosed more easily, and even overriding the ``self"-signal, when rigidified \cite{sosale_2015_cell}. The model results 
give insight into the observed dependence of engulfment and trogocytosis on cell membrane tension \cite{rollins2025target} and rigidity \cite{shen2025trogocytosis}.

Note that our model predicts that successful engulfment of soft targets is facilitated by the cell employing weaker cytoskeletal forces (Fig.\ref{fig:passive}). Altogether, these results suggest possible future interventions to enhance or inhibit phagocytosis and trogocytosis based on the interplay between the target's stiffness and the protrusive activity of the engulfing cell.

Our model is using only the most minimal physical components and forces, and therefore offers a path to obtaining deep and general understanding of an important biological process which is shared by many cell types \cite{miyake2021role,vorselen2022dynamics,barbera2025trogocytosis}. Future extensions of our modeling approach could include additional processes, such as myosin-induced contractility and the effects of actin treadmilling-induced forces, in addition to more complex description of the adhesion between the cell and its target (which may itself depend on stiffness and applied forces \cite{cornell2025target}).

\section{Methods and Material}

\textbf{GUV electroformation}---Solutions containing 0.25 mg total lipids were spread evenly on slides coated with indium tin oxide (70-100 $\Omega$/sq; Sigma Aldrich). The slides were placed under vacuum for $>$30 min to allow for complete evaporation of chloroform. A capacitor was created by sandwiching a 0.3-mm rubber septum between two lipid-coated slides. The gap was filled with ~200 $\mu$L of 300 mM sucrose (hyperosmotic solution compared to PBS (285 mOsm) to make high-tension GUVs) or 270 mM sucrose (hypoosmotic solution compared to PBS to make low-tension GUVs). Sucrose solution osmolarity was measured using an osmometer (Precision Systems). GUVs 10 to 100 $\mu$m in diameter were electroformed by application of an AC voltage of 1.5 V at 10 Hz across the capacitor for 1 h at 55$^o$C.

\begin{figure*}
    \centering    \includegraphics[width=\linewidth]{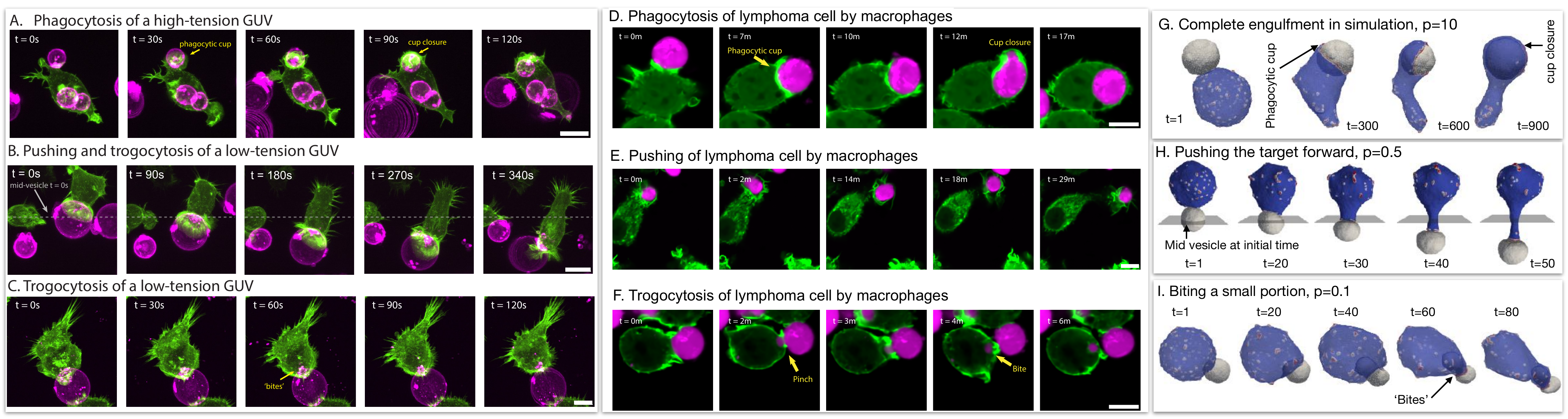}
    \caption{(A) Giant unilamellar vesicles (GUVs) composed of POPC and biotin-DOPE and filled with a hyperosmotic sucrose solution (300 mOsm), leading to taut high-tension vesicles, are opsonized with AlexaFluor647 anti-biotin and mixed with LifeAct-GFP-expressing macrophages. During phagocytosis, the macrophage forms a phagocytic cup, identifiable from enriched actin in a ring, that encircles the GUV. The macrophage fully engulfs the GUV within tens of seconds. (B) GUVs filled with a hypoosmotic sucrose solution (270 mOsm) are at low tension and can be ‘pushed’ by a macrophage, as shown by the position of the vesicle over time relative to the mid-plane at t=0s. (C). When low-tension GUVs are trogocytosed by a macrophage, punctate ‘bites’ can be observed within the macrophage. Scale bar is 5 µm. (D-F) Time-lapse confocal microscopy images show interactions between macrophages (Lifeact-mScarlet3, pseudo-colored in green) and antibody-opsonized lymphoma cells (pseudo-colored in magenta). D) Example of a macrophage engulfing an entire target cell. E) Example of a macrophage pushing the target cell. F) Example of a macrophages trogocytosing a fragment of the target cell. Scale bars are 10 $\mu$m. In simulations, (G) the complete engulfment of the target vesicle when its osmotic pressure is high $p=10~k_BT l_{\rm min}^{-3}$. (H) The target is pushed when the osmotic pressure is intermediate $p=0.5~ k_BT l_{\rm min}^{-3}$. A plane that is shown perpendicular to the pushing direction through the middle of the target vesicle's initial position. (I) For a very low osmotic pressure, $p=0.1~ k_BT l_{\rm min}^{-3}$, the cell-like vesicle takes a bite from the target.}
    \label{fig:three_behavior}
\end{figure*}

\textbf{Imaging techniques}---All live cells were maintained at 37$^o$C, 5$~\%$ $CO_2$ with a stage top incubator (Okolab) during imaging. For confocal microscopy, cells were imaged with a spinning disk confocal microscope (Eclipse Ti, Nikon) with a spinning disk (Yokogawa CSU-X, Andor), sCMOS camera (Prime 95B, Photometrics), and a 60x objective (Apo TIRF, 1.49NA, oil, Nikon). The spinning disk confocal microscope was controlled with Nikon Elements (Nikon). Images were analyzed and prepared using FIJI (imagej.net/software/fiji).

\textbf{Phagocytosis/trogocytosis of GUVs}---50,000 macrophages were seeded in wells of an 8-well glass-bottom plate (CellVis) in 100 $\mu$L of RPMI 1640 medium. Post-seeding, cells were incubated at 37$^o$C, 5$~\%$ $CO_2$ for 3-4 hours before target addition. 100 $\mu$L of low-tension GUVs or high-tension GUVs (~1 million GUVs counted with an impedance-based cell counter (Scepter, SigmaAldrich)) were prepared with 4 $\mu$M AlexaFluor647 anti-biotin IgG in PBS and allowed to incubate with gentle rotation for $>10$ minutes. After washing, GUVs were added to macrophage-seeded wells on the stage top incubator of the microscope. 

\textbf{Time-lapse confocal microscopy of antibody-dependent cellular phagocytosis} ---
J774A.1 macrophages and Raji B lymphocytes were obtained from DSMZ (ACC170, ACC319). J774A.1 that stably express Lifeact-mScarlet3 \cite{gadella2023mscarlet3} were cultured at 37$^o$C with 5$~\%$ $CO_2$ in DMEM (Wisent, 319-005) supplemented with 10$~\%$ of heat-inactivated FBS (Wisent, 090-150). Raji were cultured were cultured at 37$^o$C with 5$~\%$ $CO_2$ in RPMI-1640 (Wisent, 350-007) supplemented with 10$~\%$ of heat-inactivated FBS (Wisent, 090-150). 24 hours before imaging, 100,000 macrophages were seeded into 18 mm circular \#1.5 coverslips. Prior to the experiment, Raji cells were stained with Calcein AM viability dye (eBioscience, 65-0853-78) at 1 $\mu$M for 30 minutes. Macrophages were transferred to a Chamlide CMB imaging chamber (Live Cell Instrument) and incubated in DMEM without phenol red, containing 25 mM HEPES (Wisent, 319-066). 200,000 Raji cells were added to the macrophages, and 2 $\mu$g/ml of anti-human CD20 antibody (BioXcell, Rituximab) and 10 $\mu$g/ml of anti-human CD47 antibody (BioXcell, B6.H12) were supplemented. Images were acquired every 20 seconds for 2 hours using a 40x 1.3 NA oil immersion objective on a Nikon Eclipse Ti2-E, equipped with a Yokogawa SCU-W1 spinning disk, a Hamamatsu Orca-Fusion BT sCMOS camera, and a stage-top incubator (Tokai Hit) to maintain cells at 37$^o$C. Images were acquired using Nikon NIS Elements and analyzed using Fiji is just ImageJ 1.54p.

\textbf{Lipid coating of hydrogel beads functionalized with TDA}---
First, unilamellar vesicles (LUVs) were prepared by the extrusion method. POPS (1-palmitoyl-2-oleoyl-sn-glycero-3-phospho-L-serine, Avanti Polar Lipids), POPC (1-palmitoyl-2-oleoyl-sn-glycero-3-phosphocholine, Avanti Polar Lipids) and Texas RedTM DHPE (Texas RedTM 1,2-Dihexadecanoyl-sn-Glycero-3- Phospho-ethanolamine, Invitrogen) dissolved in chloroform were mixed in a 80:19:1 molar ratio and dried. The dried lipid cake was then hydrated in PBS to a final concentration of 1 mM lipid suspension followed by 5 freeze-thaw cycles. Then, LUVs were prepared by 11-fold extrusion through a 100-nm polycarbonate filter (Avanti Polar Lipids) using a Mini-Extruder (Avanti Polar Lipids). Next, $50~\mu$L of vesicle solution were incubated for 45 min with $5~\mu$L of functionalized hydrogel beads followed 5 freeze-thaw cycles. Unbound vesicles were washed 5 times with PBS by centrifuging the solution for 1 min at 5000 g. Coated hydrogel beads were stored at 4ºC and used up to a week. 

\textbf{Zebrafish handling and maintenance}---Zebrafish were maintained at the aquatic facility of the Parc de Recerca Biomèdica de Barcelona (PRBB) following the standard procedures proved by the Institutional Animal Care and Use Ethic Committee (PRBB–IACUEC) according to national and European regulations. All the experiments were performed following the principles of the 3Rs. Eggs were kept in E3 medium (5 mM NaCl, 0.17 mM KCl, 0.33 mM CaCl2, 0.33 mM MgSO4) at 28°C. Embryos were staged based on morphological criteria and hours post fertilization (hpf). All embryos used in this study were obtained from the Tg(actb1:Lifeact-GFP) \cite{behrndt2012forces} zebrafish line. 

\textbf{Transplantation of synthetic targets}---
Acceptor embryos were dechorionated at sphere stage (4 hpf) and placed into a custom-made transplantation agarose mold (Adaptative Science Tools) containing Danieau’s solution (58 mM NaCl, 0.7 mM KCl, 0.4 mM MgSO4, 0.6 mM Ca(NO3)2, 5 mM HEPES). Synthetic targets were transplanted into the animal cap of acceptor embryos, close to the embryonic epithelial layer, using a Celltram Vario device (Eppendorf) in combination with an ES Blastocyst-pipette needle (Biomedical Instruments, VESbv-20-0-0-55).

\textbf{Apoptosis induction of a subpopulation of cells}---
Apoptosis was induced in a subpopulation of embryonic cells as described previously \cite{hoijman2021cooperative}. In short, 2 pg of zebrafish bax mRNA \cite{popgeorgiev2011apoptotic} were co-injected with 100 pg of the membrane marker lyn-tdTomato mRNA into 16-32 cell stage embryos for a mosaic expression. mRNAs were synthetized from pCS2+ plasmids using the SP6 mMessage mMachine kit (Ambion, AM1340). 

For more details of experimental and simulation method, see SI section S14.

\section{Code availability}
The MATLAB code for analyzing confocal images and deriving particle shape is publicly available on \href{https://gitlab.com/dvorselen/DAAMparticle_Shape_Analysis}{https://gitlab.com/dvorselen/DAAMparticle\_Shape\_Analysis}. The Python code for traction force analysis is available on \href{https://gitlab.com/micronano_public/ShElastic}{https://gitlab.com/micronano\_public/ShElastic}.


\begin{acknowledgements}
N.S.G., incumbent of the Lee and William Abramowitz Professorial Chair of Biophysics, acknowledges support from the Israel Science Foundation (Grant No. 207/22) and the Harold Perlman Family. D.A.F. was supported by the National Science Foundation Center for Cellular Construction (DBI-1548297), the National Institutes of Health (R01GM134137), and the Chan Zuckerberg Biohub Investigator program. C.E.C. was supported by the James S. McDonnell Foundation Postdoctoral Fellowship. A.I. and S.P. acknowledge support from the Slovenian Research Agency (ARIS; project J3-60063 and programme P2-0232), the EU Horizon 2020 Marie Skłodowska-Curie Staff Exchange project *FarmEVs* (Grant No. 101131175), and COST Action CA22153. V.R. was supported by the Ministerio de Ciencia e Innovación (PID2020-117011GB-I00), HFSP (RGY0079/2020), the European Union's Horizon Europe programme (BREAKDANCE, Grant No. 101072123), and the Austrian Science Fund (FWF; 10.55776/PIN3977225). M.B. acknowledges support from the Ministerio de Ciencia, Innovación y Universidades and Fondo Social Europeo (FSE; PRE2020-092691).
\end{acknowledgements}

\textit{\textbf{Author contributions}}---{SS and NSG designed the study and computational model. SS developed the code and performed simulations and data analysis. CEC, DAF, MKS, VJ, YP, DV, VR, and MB performed the experimental work, with VR and MB also generating lipid-coated hydrogel beads. AI and SP developed the original single-vesicle code. SS and NSG wrote the manuscript, and all authors reviewed and edited it.
}

\bibliography{sample}

@article{Pelling2009,
  title = {Mechanical dynamics of single cells during early apoptosis},
  volume = {66},
  ISSN = {0886-1544},
  url = {http://dx.doi.org/10.1002/cm.20391},
  DOI = {10.1002/cm.20391},
  number = {7},
  journal = {Cell Motility},
  publisher = {Wiley},
  author = {Pelling,  Andrew E. and Veraitch,  Farlan S. and Chu,  Carol Pui‐Kei and Mason,  Chris and Horton,  Michael A.},
  year = {2009},
  month = June,
  pages = {409–422}
}

@article{VanderMeeren2020,
  title = {AFM Analysis Enables Differentiation between Apoptosis,  Necroptosis,  and Ferroptosis in Murine Cancer Cells},
  volume = {23},
  ISSN = {2589-0042},
  url = {http://dx.doi.org/10.1016/j.isci.2020.101816},
  DOI = {10.1016/j.isci.2020.101816},
  number = {12},
  journal = {iScience},
  publisher = {Elsevier BV},
  author = {Van der Meeren,  Louis and Verduijn,  Joost and Krysko,  Dmitri V. and Skirtach,  André G.},
  year = {2020},
  month = Dec,
  pages = {101816}
}

@article{Jaumouill2019_Nat_cell_bio,
  title = {Coupling of $\beta$2 integrins to actin by a mechanosensitive molecular clutch drives complement receptor-mediated phagocytosis},
  volume = {21},
  ISSN = {1476-4679},
  url = {http://dx.doi.org/10.1038/s41556-019-0414-2},
  DOI = {10.1038/s41556-019-0414-2},
  number = {11},
  journal = {Nature Cell Biology},
  publisher = {Springer Science and Business Media LLC},
  author = {Jaumouillé,  Valentin and Cartagena-Rivera,  Alexander X. and Waterman,  Clare M.},
  year = {2019},
  month = Oct,
  pages = {1357–1369}
}

@article{behrndt2012forces,
  title={Forces driving epithelial spreading in zebrafish gastrulation},
  author={Behrndt, Martin and Salbreux, Guillaume and Campinho, Pedro and Hauschild, Robert and Oswald, Felix and Roensch, Julia and Grill, Stephan W and Heisenberg, Carl-Philipp},
  journal={Science},
  volume={338},
  number={6104},
  pages={257--260},
  year={2012},
  publisher={American Association for the Advancement of Science}
}

@article {sosale_2015_cell,
	Title = {Cell rigidity and shape override CD47's "self"-signaling in phagocytosis by hyperactivating myosin-II},
	Author = {Sosale, Nisha G and Rouhiparkouhi, Tahereh and Bradshaw, Andrew M and Dimova, Rumiana and Lipowsky, Reinhard and Discher, Dennis E},
	DOI = {10.1182/blood-2014-06-585299},
	Number = {3},
	Volume = {125},
	Month = {January},
	Year = {2015},
	Journal = {Blood},
	ISSN = {0006-4971},
	Pages = {542—552},
	URL = {https://europepmc.org/articles/PMC4296014},
}

@article{debnath2025wrapping,
  title={Wrapping of Nano-and Microgels by Lipid-Bilayer Membranes},
  author={Debnath, Tanwi and Midya, Jiarul and Auth, Thorsten and Gompper, Gerhard},
  journal={ACS Macro Letters},
  volume={14},
  number={10},
  pages={1412--1417},
  year={2025},
  publisher={ACS Publications}
}

@article{sorokina2024computer,
  title={Computer simulations of responsive nanogels at lipid membrane},
  author={Sorokina, Anastasia S and Gumerov, Rustam A and Noguchi, Hiroshi and Potemkin, Igor I},
  journal={Macromolecular Rapid Communications},
  volume={45},
  number={21},
  pages={2400406},
  year={2024},
  publisher={Wiley Online Library}
}

@article{yi2016incorporation,
  title={Incorporation of soft particles into lipid vesicles: Effects of particle size and elasticity},
  author={Yi, Xin and Gao, Huajian},
  journal={Langmuir},
  volume={32},
  number={49},
  pages={13252--13260},
  year={2016},
  publisher={ACS Publications}
}

@article{agudo2015critical,
  title={Critical particle sizes for the engulfment of nanoparticles by membranes and vesicles with bilayer asymmetry},
  author={Agudo-Canalejo, Jaime and Lipowsky, Reinhard},
  journal={ACS nano},
  volume={9},
  number={4},
  pages={3704--3720},
  year={2015},
  publisher={ACS Publications}
}

@article{vacha2011receptor,
  title={Receptor-mediated endocytosis of nanoparticles of various shapes},
  author={V{\'a}cha, Robert and Martinez-Veracoechea, Francisco J and Frenkel, Daan},
  journal={Nano letters},
  volume={11},
  number={12},
  pages={5391--5395},
  year={2011},
  publisher={ACS Publications}
}

@article{zhang2024signaling,
  title={Signaling-biophysical modeling unravels mechanistic control of red blood cell phagocytosis by macrophages in sickle cell disease},
  author={Zhang, Yu and Qiang, Yuhao and Li, He and Li, Guansheng and Lu, Lu and Dao, Ming and Karniadakis, George E and Popel, Aleksander S and Zhao, Chen},
  journal={PNAS nexus},
  volume={3},
  number={2},
  pages={pgae031},
  year={2024},
  publisher={Oxford University Press US}
}

@article{Mali2025_nat_protocols,
  title = {Using tunable hydrogel microparticles to measure cellular forces},
  ISSN = {1750-2799},
  url = {http://dx.doi.org/10.1038/s41596-025-01281-2},
  DOI = {10.1038/s41596-025-01281-2},
  journal = {Nature Protocols},
  publisher = {Springer Science and Business Media LLC},
  author = {Mali,  Alvja and Peeters,  Youri and Rodrigues de Mercado,  Rick and Settle,  Alexander H. and Footer,  Matthew J. and Srinivas,  Mangala and Theriot,  Julie A. and Vorselen,  Daan},
  year = {2025},
  month = dec 
}

@article{Vorselen2020_nat_comm,
  title = {Microparticle traction force microscopy reveals subcellular force exertion patterns in immune cell–target interactions},
  volume = {11},
  ISSN = {2041-1723},
  url = {http://dx.doi.org/10.1038/s41467-019-13804-z},
  DOI = {10.1038/s41467-019-13804-z},
  number = {1},
  journal = {Nature Communications},
  publisher = {Springer Science and Business Media LLC},
  author = {Vorselen,  Daan and Wang,  Yifan and de Jesus,  Miguel M. and Shah,  Pavak K. and Footer,  Matthew J. and Huse,  Morgan and Cai,  Wei and Theriot,  Julie A.},
  year = {2020},
  month = jan 
}

@article{barbera2025trogocytosis,
  title={Trogocytosis of chimeric antigen receptors between T cells is regulated by their transmembrane domains},
  author={Barbera, Stefano and Schuiling, Matthijs JA and Sanjaya, Nathaniel A and Pietil{\"a}, Ilkka and Sar{\'e}n, Tina and Essand, Magnus and Dimberg, Anna},
  journal={Science immunology},
  volume={10},
  number={103},
  pages={eado2054},
  year={2025},
  publisher={American Association for the Advancement of Science}
}

@article{champion2006role,
  title={Role of target geometry in phagocytosis},
  author={Champion, Julie A and Mitragotri, Samir},
  journal={Proceedings of the National Academy of Sciences},
  volume={103},
  number={13},
  pages={4930--4934},
  year={2006},
  publisher={National Academy of Sciences}
}

@article{miyake2021role,
  title={The role of trogocytosis in the modulation of immune cell functions},
  author={Miyake, Kensuke and Karasuyama, Hajime},
  journal={Cells},
  volume={10},
  number={5},
  pages={1255},
  year={2021}
}

@article{vorselen2022dynamics,
  title={Dynamics of phagocytosis mediated by phosphatidylserine},
  author={Vorselen, Daan},
  journal={Biochemical Society Transactions},
  volume={50},
  number={5},
  pages={1281--1291},
  year={2022},
  publisher={Portland Press Ltd.}
}

@article{Vorselen2021_elife,
  title = {Phagocytic ‘teeth’ and myosin-II ‘jaw’ power target constriction during phagocytosis},
  volume = {10},
  ISSN = {2050-084X},
  url = {http://dx.doi.org/10.7554/eLife.68627},
  DOI = {10.7554/elife.68627},
  journal = {eLife},
  publisher = {eLife Sciences Publications,  Ltd},
  author = {Vorselen,  Daan and Barger,  Sarah R and Wang,  Yifan and Cai,  Wei and Theriot,  Julie A and Gauthier,  Nils C and Krendel,  Mira},
  year = {2021},
  month = oct 
}

@misc{Daan_2021_figshare,
  doi = {10.6084/M9.FIGSHARE.16666864},
  url = {https://figshare.com/articles/dataset/Phagocytic_microscopy_and_MP-TFM_assay_with_RAW_macrophages_upon_treatment_with_cytoskeletal_inhibitors/16666864},
  author = {Vorselen,  Daan and Barger,  Sarah and Theriot,  Julie and Gauthier,  Nils and Krendel,  Mira},
  title = {Phagocytic microscopy and MP-TFM assay with RAW macrophages upon treatment with cytoskeletal inhibitors},
  publisher = {figshare},
  year = {2021},
  copyright = {Creative Commons Attribution 4.0 International}
}

@article{frank2008prevention,
  title={Prevention of microvesiculation by adhesion of buds to the mother cell membrane—a possible anticoagulant effect of healthy donor plasma},
  author={Frank, Mojca and Man{\v{c}}ek-Keber, Mateja and Kr{\v{z}}an, Mojca and Sodin-{\v{S}}emrl, Sne{\v{z}}na and Jerala, Roman and Igli{\v{c}}, Ale{\v{s}} and Rozman, Bla{\v{z}} and Kralj-Igli{\v{c}}, Veronika},
  journal={Autoimmunity Reviews},
  volume={7},
  number={3},
  pages={240--245},
  year={2008},
  publisher={Elsevier}
}

@article{urbanija2008attachment,
  title={Attachment of $\beta$2-glycoprotein I to negatively charged liposomes may prevent the release of daughter vesicles from the parent membrane},
  author={Urbanija, Jasna and Babnik, Bla{\v{z}} and Frank, Mojca and Tom{\v{s}}i{\v{c}}, Nejc and Rozman, Bla{\v{z}} and Kralj-Igli{\v{c}}, Veronika and Igli{\v{c}}, Ale{\v{s}}},
  journal={European Biophysics Journal},
  volume={37},
  number={7},
  pages={1085--1095},
  year={2008},
  publisher={Springer}
}

@article{pavlivc2009encapsulation,
  title={Encapsulation of small spherical liposome into larger flaccid liposome induced by human plasma proteins},
  author={Pavli{\v{c}}, JI and Mare{\v{s}}, T and Be{\v{s}}ter, J and Jan{\v{s}}a, V and Daniel, M and Igli{\v{c}}, A},
  journal={Computer Methods in Biomechanics and Biomedical Engineering},
  volume={12},
  number={2},
  pages={147--150},
  year={2009},
  publisher={Taylor \& Francis}
}

@article{jaumouille2020physical,
  title={Physical constraints and forces involved in phagocytosis},
  author={Jaumouill{\'e}, Valentin and Waterman, Clare M},
  journal={Frontiers in immunology},
  volume={11},
  pages={1097},
  year={2020},
  publisher={Frontiers Media SA}
}

@article{vorselen2020mechanical,
  title={A mechanical perspective on phagocytic cup formation},
  author={Vorselen, Daan and Labitigan, Ramon Lorenzo D and Theriot, Julie A},
  journal={Current opinion in cell biology},
  volume={66},
  pages={112--122},
  year={2020},
  publisher={Elsevier}
}

@article{swanson2008shaping,
  title={Shaping cups into phagosomes and macropinosomes},
  author={Swanson, Joel A},
  journal={Nature reviews Molecular cell biology},
  volume={9},
  number={8},
  pages={639--649},
  year={2008},
  publisher={Nature Publishing Group UK London}
}

@article{mylvaganam2021cytoskeleton,
  title={The cytoskeleton in phagocytosis and macropinocytosis},
  author={Mylvaganam, Sivakami and Freeman, Spencer A and Grinstein, Sergio},
  journal={Current Biology},
  volume={31},
  number={10},
  pages={R619--R632},
  year={2021},
  publisher={Elsevier}
}

@incollection{PANAH2026651,
title = {Signaling of Phagocytosis},
editor = {Paul M. Kaye},
booktitle = {Encyclopedia of Immunobiology (Second Edition)},
publisher = {Academic Press},
edition = {Second Edition},
address = {Oxford},
pages = {651-669},
year = {2026},
isbn = {978-0-12-824480-7},
doi = {https://doi.org/10.1016/B978-0-12-824465-4.00199-X},
url = {https://www.sciencedirect.com/science/article/pii/B978012824465400199X},
author = {Abdolrasol K Panah and Valentin Jaumouillé},
}

@article{uribe2020phagocytosis,
  title={Phagocytosis: our current understanding of a universal biological process},
  author={Uribe-Querol, Eileen and Rosales, Carlos},
  journal={Frontiers in immunology},
  volume={11},
  pages={1066},
  year={2020},
  publisher={Frontiers Media SA}
}

@article{flannagan2012cell,
  title={The cell biology of phagocytosis},
  author={Flannagan, Ronald S and Jaumouill{\'e}, Valentin and Grinstein, Sergio},
  journal={Annual Review of Pathology: Mechanisms of Disease},
  volume={7},
  number={1},
  pages={61--98},
  year={2012},
  publisher={Annual Reviews}
}

@article{gadella2023mscarlet3,
  title={mScarlet3: a brilliant and fast-maturing red fluorescent protein},
  author={Gadella Jr, Theodorus WJ and Van Weeren, Laura and Stouthamer, Jente and Hink, Mark A and Wolters, Anouk HG and Giepmans, Ben NG and Aumonier, Sylvain and Dupuy, J{\'e}r{\^o}me and Royant, Antoine},
  journal={Nature methods},
  volume={20},
  number={4},
  pages={541--545},
  year={2023},
  publisher={Nature Publishing Group US New York}
}

@article{beningo2002fc,
  title={Fc-receptor-mediated phagocytosis is regulated by mechanical properties of the target},
  author={Beningo, Karen A and Wang, Yu-li},
  journal={Journal of cell science},
  volume={115},
  number={4},
  pages={849--856},
  year={2002},
  publisher={Company of Biologists}
}

@article{lekka2016discrimination,
  title={Discrimination between normal and cancerous cells using AFM},
  author={Lekka, Ma{\l}gorzata},
  journal={Bionanoscience},
  volume={6},
  number={1},
  pages={65--80},
  year={2016},
  publisher={Springer}
}

@article{yang2025single,
  title={Single-cell parallel plate mechanics by side-view optical microscopy-assisted atomic force microscopy},
  author={Yang, Mingyang and Yang, Yanqi and Liu, Lianqing and Li, Mi},
  journal={Nanoscale Advances},
  volume={7},
  number={8},
  pages={2158--2165},
  year={2025},
  publisher={Royal Society of Chemistry}
}

@article{sadhu2024minimal,
  title={A minimal physical model for curvotaxis driven by curved protein complexes at the cell’s leading edge},
  author={Sadhu, Raj Kumar and Luciano, Marine and Xi, Wang and Martinez-Torres, Cristina and Schr{\"o}der, Marcel and Blum, Christoph and Tarantola, Marco and Villa, Stefano and Peni{\v{c}}, Samo and Igli{\v{c}}, Ale{\v{s}} and others},
  journal={Proceedings of the National Academy of Sciences},
  volume={121},
  number={12},
  pages={e2306818121},
  year={2024},
  publisher={National Academy of Sciences}
}

@article{Sadhukhan2025,
  title = {Modeling how lamellipodia-driven cells maintain persistent migration and interact with external barriers},
  author = {Sadhukhan,  Shubhadeep and Martinez-Torres,  Cristina and Penič,  Samo and Beta,  Carsten and Iglič,  Aleš and Gov,  Nir},
  journal = {Phys. Rev. Res.},
  volume = {7},
  issue = {1},
  pages = {013319},
  numpages = {13},
  year = {2025},
  month = {Mar},
  publisher = {American Physical Society},
  doi = {10.1103/PhysRevResearch.7.013319},
  url = {https://link.aps.org/doi/10.1103/PhysRevResearch.7.013319}
}

@article{tollis2010zipper,
  title={The zipper mechanism in phagocytosis: energetic requirements and variability in phagocytic cup shape},
  author={Tollis, Sylvain and Dart, Anna E and Tzircotis, George and Endres, Robert G},
  journal={BMC systems biology},
  volume={4},
  number={1},
  pages={149},
  year={2010},
  publisher={Springer}
}

@article{wu2025wave,
  title={WAVE complex forms linear arrays at negative membrane curvature to instruct lamellipodia formation},
  author={Wu, Muziyue and Sadhu, Raj Kumar and Meyer, Kirstin and Tang, Ziqi and Marchando, Paul and Woolfson, Derek N and Gov, Nir S and Weiner, Orion D},
  journal={Journal of Cell Biology},
  volume={224},
  number={9},
  pages={e202410098},
  year={2025},
  publisher={Rockefeller University Press}
}

@article{pipathsouk2021wave,
  title={The WAVE complex associates with sites of saddle membrane curvature},
  author={Pipathsouk, Anne and Brunetti, Rachel M and Town, Jason P and Graziano, Brian R and Breuer, Artu and Pellett, Patrina A and Marchuk, Kyle and Tran, Ngoc-Han T and Krummel, Matthew F and Stamou, Dimitrios and others},
  journal={Journal of Cell Biology},
  volume={220},
  number={8},
  pages={e202003086},
  year={2021},
  publisher={Rockefeller University Press}
}

@article{begemann2019mechanochemical,
  title={Mechanochemical self-organization determines search pattern in migratory cells},
  author={Begemann, Isabell and Saha, Tanumoy and Lamparter, L and Rathmann, I and Grill, D and Golbach, L and Rasch, C and Keller, U and Trappmann, B and Matis, M and others},
  journal={Nature Physics},
  volume={15},
  number={8},
  pages={848--857},
  year={2019},
  publisher={Nature Publishing Group UK London}
}

@article{scita2008irsp53,
  title={IRSp53: crossing the road of membrane and actin dynamics in the formation of membrane protrusions},
  author={Scita, Giorgio and Confalonieri, Stefano and Lappalainen, Pekka and Suetsugu, Shiro},
  journal={Trends in cell biology},
  volume={18},
  number={2},
  pages={52--60},
  year={2008},
  publisher={Elsevier}
}

@article{linkner2014inverse,
  title={The inverse BAR domain protein IBARa drives membrane remodeling to control osmoregulation, phagocytosis and cytokinesis},
  author={Linkner, Joern and Witte, Gregor and Zhao, Hongxia and Junemann, Alexander and Nordholz, Benjamin and Runge-Wollmann, Petra and Lappalainen, Pekka and Faix, Jan},
  journal={Journal of Cell Science},
  volume={127},
  number={6},
  pages={1279--1292},
  year={2014},
  publisher={The Company of Biologists Bidder Building, 140 Cowley Road, Cambridge, CB4~…}
}

@article{vorselen2020microparticle,
  title={Microparticle traction force microscopy reveals subcellular force exertion patterns in immune cell--target interactions},
  author={Vorselen, Daan and Wang, Yifan and de Jesus, Miguel M and Shah, Pavak K and Footer, Matthew J and Huse, Morgan and Cai, Wei and Theriot, Julie A},
  journal={Nature communications},
  volume={11},
  number={1},
  pages={20},
  year={2020},
  publisher={Nature Publishing Group UK London}
}

@article{settle2024beta2,
  title={$\beta$2 integrins impose a mechanical checkpoint on macrophage phagocytosis},
  author={Settle, Alexander H and Winer, Benjamin Y and de Jesus, Miguel M and Seeman, Lauren and Wang, Zhaoquan and Chan, Eric and Romin, Yevgeniy and Li, Zhuoning and Miele, Matthew M and Hendrickson, Ronald C and others},
  journal={Nature Communications},
  volume={15},
  number={1},
  pages={8182},
  year={2024},
  publisher={Nature Publishing Group UK London}
}

@article{torres2022interacting,
  title={Interacting active surfaces: A model for three-dimensional cell aggregates},
  author={Torres-S{\'a}nchez, Alejandro and Kerr Winter, Max and Salbreux, Guillaume},
  journal={PLOS Computational Biology},
  volume={18},
  number={12},
  pages={e1010762},
  year={2022},
  publisher={Public Library of Science San Francisco, CA USA}
}

@article{shen2025trogocytosis,
  title={Trogocytosis: revealing new insights into parasite--host interactions},
  author={Shen, Jia and Li, Xuesong and Hide, Geoff and Lun, Zhao-Rong and Wu, Zhongdao},
  journal={Trends in Parasitology},
  year={2025},
  publisher={Elsevier}
}

@article{rollins2025target,
  title={Target cell adhesion limits macrophage phagocytosis and promotes trogocytosis},
  author={Rollins, Kirstin R and Fiaz, Sareen and Datta, Ishwaree and Morrissey, Meghan A},
  journal={Journal of Cell Biology},
  volume={224},
  number={11},
  pages={e202502034},
  year={2025},
  publisher={Rockefeller University Press}
}

@article{meijering2012methods,
  title={Methods for cell and particle tracking},
  author={Meijering, Erik and Dzyubachyk, Oleh and Smal, Ihor},
  journal={Methods in enzymology},
  volume={504},
  pages={183--200},
  year={2012},
  publisher={Elsevier}
}

@article{popgeorgiev2011apoptotic,
  title={The apoptotic regulator Nrz controls cytoskeletal dynamics via the regulation of Ca2+ trafficking in the zebrafish blastula},
  author={Popgeorgiev, Nikolay and Bonneau, Benjamin and Ferri, Karine F and Prudent, Julien and Thibaut, Julien and Gillet, Germain},
  journal={Developmental cell},
  volume={20},
  number={5},
  pages={663--676},
  year={2011},
  publisher={Elsevier}
}

@article{hoijman2021cooperative,
  title={Cooperative epithelial phagocytosis enables error correction in the early embryo},
  author={Hoijman, Esteban and H{\"a}kkinen, Hanna-Maria and Tolosa-Ramon, Queralt and Jimenez-Delgado, Senda and Wyatt, Chris and Miret-Cuesta, Marta and Irimia, Manuel and Callan-Jones, Andrew and Wieser, Stefan and Ruprecht, Verena},
  journal={Nature},
  volume={590},
  number={7847},
  pages={618--623},
  year={2021},
  publisher={Nature Publishing Group UK London}
}

@article{marevs2012determination,
  title={Determination of the strength of adhesion between lipid vesicles},
  author={Mare{\v{s}}, Tom{\'a}{\v{s}} and Daniel, Matej and Igli{\v{c}}, Ale{\v{s}} and Kralj-Igli{\v{c}}, Veronika and Fo{\v{s}}nari{\v{c}}, Miha},
  journal={The Scientific World Journal},
  volume={2012},
  number={1},
  pages={146804},
  year={2012},
  publisher={Wiley Online Library}
}

@article{tordeux2002analytical,
  title={Analytical characterization of adhering vesicles},
  author={Tordeux, C and Fournier, J-B and Galatola, P},
  journal={Physical Review E},
  volume={65},
  number={4},
  pages={041912},
  year={2002},
  publisher={APS}
}

@article{sadhu2021modelling,
  title={Modelling cellular spreading and emergence of motility in the presence of curved membrane proteins and active cytoskeleton forces},
  author={Sadhu, Raj Kumar and Peni{\v{c}}, Samo and Igli{\v{c}}, Ale{\v{s}} and Gov, Nir S},
  journal={The European Physical Journal Plus},
  volume={136},
  number={5},
  pages={495},
  year={2021},
  publisher={Springer Berlin Heidelberg}
}

@article{Fonari2019,
  doi = {10.1039/c8sm02356e},
  url = {https://doi.org/10.1039/c8sm02356e},
  year = {2019},
  publisher = {Royal Society of Chemistry ({RSC})},
  volume = {15},
  number = {26},
  pages = {5319--5330},
  author = {Miha Fo{\v{s}}nari{\v{c}} and Samo Peni{\v{c}} and Ale{\v{s}} Igli{\v{c}} and Veronika Kralj-Igli{\v{c}} and Mitja Drab and Nir S. Gov},
  title = {Theoretical study of vesicle shapes driven by coupling curved proteins and active cytoskeletal forces},
  journal = {Soft Matter}
}

@article{motahari2017actin,
  title={Actin Based Pulling Forces in Endocytosis},
  author={Motahari, Fowad and Carlsson, Anders},
  journal={Biophysical Journal},
  volume={112},
  number={3},
  pages={561a--562a},
  year={2017},
  publisher={Elsevier}
}

@article{herron2022actin,
  title={Actin nano-architecture of phagocytic podosomes},
  author={Herron, J Cody and Hu, Shiqiong and Watanabe, Takashi and Nogueira, Ana T and Liu, Bei and Kern, Megan E and Aaron, Jesse and Taylor, Aaron and Pablo, Michael and Chew, Teng-Leong and others},
  journal={Nature communications},
  volume={13},
  number={1},
  pages={4363},
  year={2022},
  publisher={Nature Publishing Group UK London}
}

@article{sopelniak2025phagocytic,
  title={Phagocytic podosomes enable efficient uptake of Candida auris by primary human macrophages},
  author={Sopelniak, Konstantin and Batlouni, Rawan and Sun, Qi-fan and Cervero, Pasquale and Linder, Stefan},
  journal={bioRxiv},
  pages={2025--09},
  year={2025},
  publisher={Cold Spring Harbor Laboratory}
}

@article{cornell2025target,
  title={Target cell cortical tension regulates macrophage trogocytosis},
  author={Cornell, Caitlin E and Chorlay, Aymeric and Krishnamurthy, Deepak and Martin, Nicholas R and Baldauf, Lucia and Fletcher, Daniel A},
  journal={Nature Cell Biology},
  pages={1--11},
  year={2025},
  publisher={Nature Publishing Group UK London}
}

@article{sadhu2023minimal,
  title={A minimal cell model for lamellipodia-based cellular dynamics and migration},
  author={Sadhu, Raj Kumar and Igli{\v{c}}, Ale{\v{s}} and Gov, Nir S},
  journal={Journal of Cell Science},
  volume={136},
  number={14},
  pages={jcs260744},
  year={2023},
  publisher={The Company of Biologists Ltd}
}

@article{rianna2020direct,
  title={Direct evidence that tumor cells soften when navigating confined spaces},
  author={Rianna, Carmela and Radmacher, Manfred and Kumar, Sanjay},
  journal={Molecular biology of the cell},
  volume={31},
  number={16},
  pages={1726--1734},
  year={2020},
  publisher={The American Society for Cell Biology}
}

@article{sadhu2023theoretical,
  title={A theoretical model of efficient phagocytosis driven by curved membrane proteins and active cytoskeleton forces},
  author={Sadhu, Raj Kumar and Barger, Sarah R and Peni{\v{c}}, Samo and Igli{\v{c}}, Ale{\v{s}} and Krendel, Mira and Gauthier, Nils C and Gov, Nir S},
  journal={Soft Matter},
  volume={19},
  number={1},
  pages={31--43},
  year={2023},
  publisher={Royal Society of Chemistry}
}

\end{document}



\title{Supplementary material:  From biting to engulfment: Target mechanics determines modes of phagocytosis through curvature--actin coupling}


\author{Shubhadeep Sadhukhan}
\email{shubhadeep.sadhukhan@weizmann.ac.il}
 \affiliation{Department of Chemical and Biological Physics, Weizmann Institute of Science, Rehovot, Israel}
 
\author{Caitlin E. Cornell}
\affiliation{
Department of Bioengineering, University of California Berkeley; Berkeley, CA USA
}
\author{Mansehaj Kaur Sandhu}
\affiliation{%
Department of Molecular Biology and Biochemistry, Simon Fraser University, Burnaby BC, Canada
}%
\author{Marta Batet Palau}
\affiliation{%
Center for Genomic Regulation (CRG), The Barcelona Institute of Science and Technology, Barcelona, Spain
}%

\author{Youri Peeters}
\affiliation{%
Department of Cell Biology and Immunology, Wageningen University and Research, Wageningen, the Netherlands
}%

\author{Stijn Hanssen}
\affiliation{%
Department of Cell Biology and Immunology, Wageningen University and Research, Wageningen, the Netherlands
}%
\author{Samo Peni\v{c}}
\affiliation{Laboratory of Physics, Faculty of Electrical Engineering, University of Ljubljana, Ljubljana, Slovenia}
\author{Ale\v{s} Igli\v{c}}
\affiliation{Laboratory of Physics, Faculty of Electrical Engineering, University of Ljubljana, Ljubljana, Slovenia}
\author{Daan Vorselen}
\affiliation{%
Department of Cell Biology and Immunology, Wageningen University and Research, Wageningen, the Netherlands
}%
\author{Daniel A. Fletcher}
\affiliation{
Department of Bioengineering, University of California Berkeley; Berkeley, CA USA
}
\affiliation{University of California Berkeley/University of California San Francisco Graduate Group in Bioengineering, CA USA}
\affiliation{Division of Biological Systems and Engineering, Lawrence Berkeley National Laboratory; Berkeley CA USA}
\affiliation{Chan Zuckerberg Biohub; San Francisco CA USA}
\author{Valentin Jaumouill\'{e}} 
\affiliation{%
Department of Molecular Biology and Biochemistry, Simon Fraser University, Burnaby BC, Canada
}%
\author{Verena Ruprecht}
\affiliation{%
Centre for Genomic Regulation (CRG), The Barcelona Institute of Science and Technology, 08003 Barcelona, Spain
}%
\affiliation{ University of Innsbruck, 6020 Innsbruck, Austria}
\author{Nir S. Gov}
\email{nir.gov@weizmann.ac.il}
 \affiliation{Department of Chemical and Biological Physics, Weizmann Institute of Science, Rehovot, Israel}
\affiliation{Department of Physiology, Development and Neuroscience, Downing Site, University of Cambridge, Cambridge, UK}

\date{\today}

\maketitle
\section{Theoretical model}\label{SI_sec:model}

We modelled the cell membrane as a three-dimensional vesicle which is described by a closed surface with $N$ vertices connected to its neighbours by bonds and it forms a dynamically triangulated, self-avoiding network, with the topology of sphere~\cite{sadhu2021modelling,sadhu2023minimal}.  $\boldsymbol{r_i}$ is the position vector of the $i$th vertex. All the lengths are measured in a scale of $l_{\rm min}$.
There is a percentage $\rho=100N_c/N$ of vertex sites that represent the curved membrane protein complexes (CMC), that induce cytoskeletal active forces. The vesicle energy has four components: The bending energy is given by,
\begin{equation}
    W_b=\frac{\kappa}{2}\int_A (C_1+C_2-C_0)^2~dA,
\end{equation}
where, $\kappa$ is the bending rigidity, $C_1$, $C_2$ are the principal curvatures and $C_0$ is the spontaneous curvature.  We consider the spontaneous curvature  $C_0=1 l_{\rm min}^{-1}$  for the CMC sites, represented in red and blue represents the bare membrane for which $C_0=0$. The protein-protein interaction energy is given by,
\begin{equation}
    W_d=-w\sum_{i<j}\mathcal{H}(r_0-r_{ij})
\end{equation}
where $\mathcal{H}$ is the Heaviside function, $r_{ij}=|\boldsymbol{r_i}-\boldsymbol{r_j}|$ is the distance between protein sites, $r_0$ is the range interaction and $w$ is the strength. We set the parameter $w=1 k_BT$ throughout the paper. The energy due to the active force is given by,
\begin{equation}
    W_F=-F\sum_i \hat{n_i}\cdot \boldsymbol{r_i}
    \label{eq:active_energy}
\end{equation}
where $F$ is the magnitude of the active force, $\hat{n_i}$ is the outward unit normal vector of the $i$th protein site vertex and $\boldsymbol{r_i}$ is the position vector of the protein. 

Finally, the adhesion energy due to an adhesive rigid substrate is given by,
\begin{equation}
    W_A=-\sum_i E_{ad}
    \label{eq:adhesion_energy}
\end{equation}
where $E_{ad}$ is the adhesion strength, and the sum runs over all the adhered vertices to the substrate i.e., the vertices that are within a range of $1l_{min}$.



\section{Detection of other vesicles}
\begin{figure}
    \centering
    \includegraphics[width=0.5\linewidth]{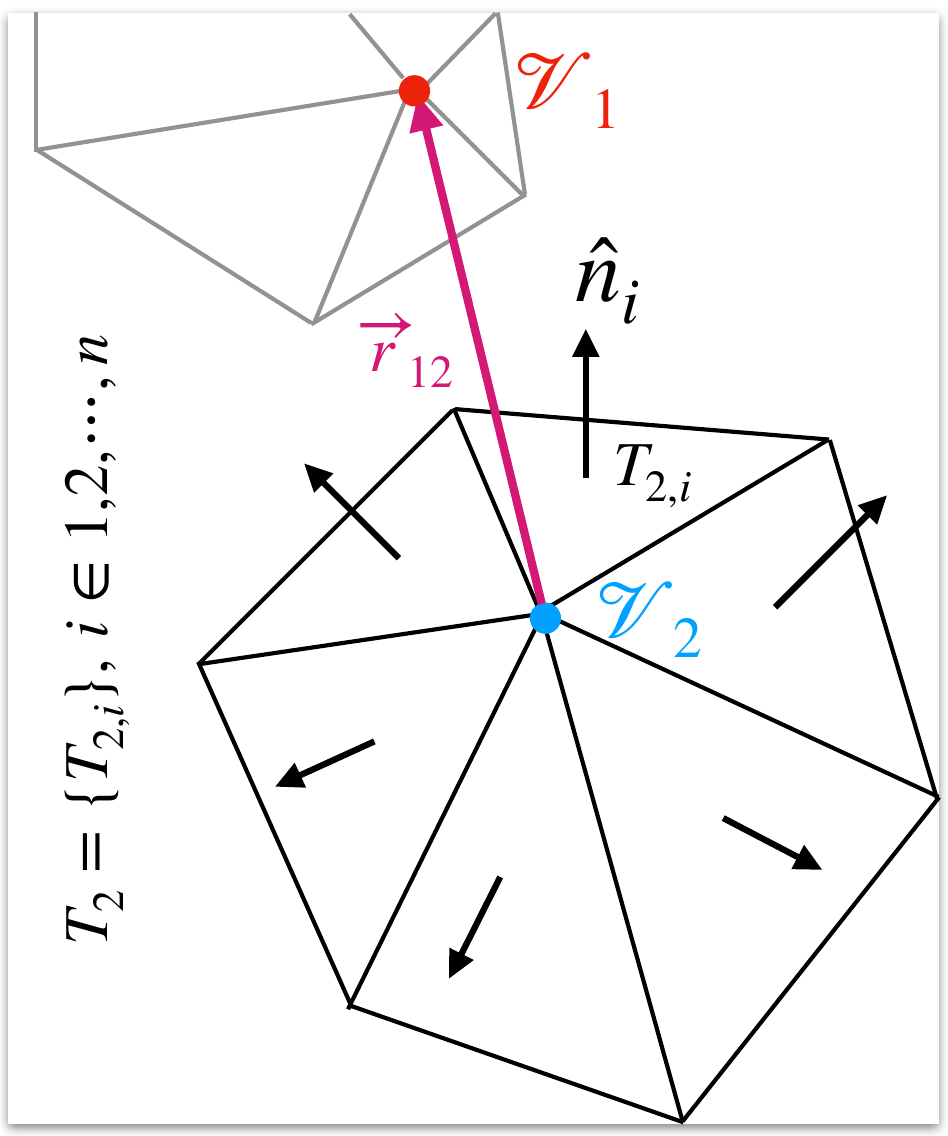}
    \caption{Demonstration of the detection of an overlap between two vesicles. Two vertices $\mathcal{V}_1$ and $\mathcal{V}_2$ belong to two different vesicles separated by the position vector $\protect\overrightarrow{r}_{12}$. We check a MC change to the position of the vertex $\mathcal{V}_1$, given that the vector $\protect\overrightarrow{r}_{12}$ magnitude is less than the minimum length scale $l_{\rm min}$. Keeping the vertex $\mathcal{V}_2$ at the center, the neighbouring triangles create a triangle list denoted by the set $T_2$. We check the dot product of the normals of the member triangles with the position vector $\protect\overrightarrow{r}_{12}$. If any of the dot products is negative, then the two vesicles are overlapping.}
    \label{fig:demo_cut}
\end{figure}

We always started the vesicles are well separated in space, i.e., not intersecting each other. In multi-vesicle simulation, it is important to make sure that one vertex of a vesicle is not going into the other vesicles during the vertex move. If we pick a vertex from a vesicle that is very close to another vertex that belongs to other vesicle, then we check whether the vesicle is cutting the other vesicle. We need to check any such pair of vertices from different vesicles. 
Let the vertex of interest $\mathcal{V}_1$ (with which we want to make a Monte-Carlo (MC) vertex movement step), is very close to another vertex $\mathcal{V}_2$, then we exploit the information of tristar associated with the vertex $\mathcal{V}_2$. We have ${T}_2$ as the "tristar" (a list of triangles, see Fig.~\ref{fig:demo_cut}) associated with $\mathcal{V}_2$. The $i$th triangle of the tristar $T_2$ is presented by $T_{2i}$. Let $\overrightarrow{r_1}$ and $\overrightarrow{r_2}$ be the position vectors of $\mathcal{V}_1$ and $\mathcal{V}_2$ respectively. We calculate the projection of the relative position of $\mathcal{V}_1$ with respect to $\mathcal{V}_2$ that is $\overrightarrow{r_{12}}=\overrightarrow{r_1}-\overrightarrow{r_2}$ on the normals $\hat{n}_i$ of each triangles $T_{2i}$ as follows:
\begin{equation}
    p_i=\overrightarrow{r}_{12}\cdot \hat{n}_i=(\overrightarrow{r_1}-\overrightarrow{r_2})\cdot\hat{n}_i
\end{equation}
where $i$ runs through all the triangles in the tristar list $T_2$ of vertex $\mathcal{V}_2$. If any $p_i$ becomes negative, it means the vertex $\mathcal{V}_1$ intersected the triangle $T_{2i}$. Hence, we abort that vertex movement.

\section{Vesicle-vesicle interaction}
Two vesicles can interact with each other through adhesion and impart the active force due to the actin-cytoskeleton.

\textbf{Vesicle-vesicle adhesion}---Two different vesicles can adhere to each other. Let $E_{\rm ad}$ be the adhesive energy per node for cell-cell adhesion. The interaction between two vesicles through adhesion becomes effective only when the two vesicles come very close and within a finite range of adhesion $1 l_{\rm min}$. Therefore, if a vertex of vesicle 1 comes within the range of cell-cell adhesion of any of the vertices belonging to another vesicle (say vesicle 2), then we count $E_{\rm ad}$ as the cell-cell adhesion energy in the energy calculation of the vertex of interest that belongs to the vesicle 1 (See Fig.1D).

\textbf{Force applied on the other vesicle}---
A vesicle can impart active force on the other vesicle. If a vertex is within the range of interaction (The interaction range is set to 1 $l_{\rm min}$) of other protein vertices belonging to the other vesicle, a vectorially added effective force is applied on the former vertex. If a set $\mathcal{S}$  of vertices belongs to another vesicle are within the interaction range of a vertex of interest $\mathcal{V}_i$ then the force applied on the vertex $\mathcal{V}_i$ is given by,
\begin{equation}
    \overrightarrow{F}^{\rm int}=\sum_{V_i\in \mathcal{S}}  \overrightarrow{F}_{V_i}.
\end{equation}


The interaction between the vesicles is mediated by the active forces applied by the others vesicles. Any vertex feels the active force due to the other vesicle's active protein sites. The effect of interaction force is included by adding the energy cost due to the interaction force given by,
\begin{equation}
    W_{\rm int}=\overrightarrow{F}^{\rm int}\cdot \overrightarrow{dr}.
\end{equation}

where, $\overrightarrow{dr}$ is the monte carlo movement of the vertex of interest.

\section{Calculation of adhered area fraction of the target vesicle}
To calculate the adhered area fraction of the target vesicle, we used a simple method by counting the number of nodes that are in the vicinity of the nodes of other vesicles and dividing it by the total number of nodes. We calculate it for different cases as follows:\\
\textbf{Adhesive area fraction calculation for the triangulated vesicle}---
To calculate the adhesion area fraction, we scan through all the vertices of a vesicle to check if the vertex is within the adhesion range from any other vesicle or the adhesive substrate. If the vertex is within the adhesion range to any other adhesive surface, then we count it as an adhered vertex. Finally, we find the approximate adhesive area fraction by dividing the number of adhered vertices $N_{\rm ad}$ by the total number of vertices $N$.
\\
\textbf{Adhesion fraction calculation for rigid sphere}---
We do not have the information for the vertices for the case of a rigid sphere. To make a similar calculation to the calculation used in the vesicle with a triangulated membrane, we generated 847 vertex points (same as the number of vertices on the target vesicle) on the sphere using a Fibonacci lattice method.

Let's say, we shall place $N^{T}$ number of vertices on a sphere of radius $R$, centred at $(x_0, ~y_0, ~z_0)$ using the Fibonacci lattice. We shall generate $N^T$ points given by,
\begin{eqnarray}
x_i=x_0+R\bigg(1-\frac{2(i-1)}{(N^T-1)}\bigg)\\\nonumber
y_i=y_0+\sqrt{(R^2-x_i^2)}\sin\phi\\\nonumber  
z_i= z_0+\sqrt{(R^2-x_i^2)}\cos \phi
\end{eqnarray}
where, $\phi=\pi(3-\sqrt{5})$ is the golden angle and the integer index $i$ runs from $1$ to $N^T$. We placed $847$ points on a sphere of radius $10~l_{\rm min}$. Then, we do the same calculation as before.

\section{Verification of radius of curvature near the contact with a single vesicle}
\begin{figure}
    \centering
    \includegraphics[width=1\linewidth]{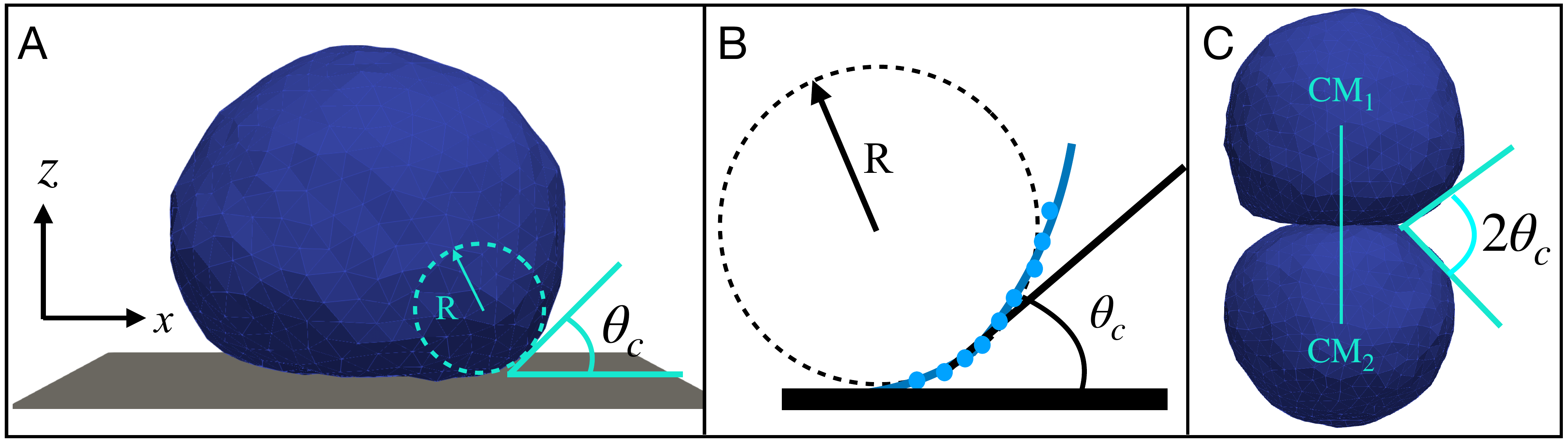}
    \caption{(A) A vesicle is adhered on an adhesive substrate. We find the cross section of the vesicle on the $x-z$ plane. Near the adhesion region we find the radius of curvature $R$. We find the angle of contact by making a linear fit with the points near the contact line: (B) Here, the relevant points along the vesicle surface are shown in blue dots. They are used to make a circular fit to find $R$, and the linear fit to find the contact angle $\theta_c$. (C) For two vesicles adhering to each other, first we join their centers of mass (${\rm CM}_1$ and ${\rm CM}_2$) and align it along $z$ direction. Then we applied the same procedure to find $R$ and $\theta_c$ as in (B). As the system has rotational symmetry around the $z$-direction, we averaged over 10 cases by rotating the system in steps of $36^o$ around the $z$ direction.}
    \label{fig:schematic_vesicle_adhesion}
\end{figure}

A single vesicle is allowed to spread on a flat adhesive substrate (Fig.~\ref{fig:schematic_vesicle_adhesion}A). After getting to the steady state, we calculated the radius of curvature near the contact when the area and the volume of the vesicle are conserved. It can be theoretically estimated from the energy terms if there are no curved proteins and no active forces. Theoretically, the radius of curvature is given by $R=\sqrt{\kappa/2E_{\rm ad}}$ \cite{marevs2012determination}. We analyzed a two-dimensional cross-section of the vesicle shape, let's say $x-z$ plane as shown in Fig.~\ref{fig:schematic_vesicle_adhesion}A. Then, we approximately identified the contact region and then fit a circle of radius $R$ that gives the radius of curvature near the contact point, as shown in Fig.~\ref{fig:schematic_vesicle_adhesion}B. Since the numerical results may fluctuate, we have found the value of $R$ by taking an average of the fit after rotating the vesicle about the $z$ axis as it has rotational symmetry about the $z$ axis. We took ten realizations by rotating the vesicle in steps of $36^o$. We can estimate an error by taking the statistical error bar. 

One can find the contact angle $\theta_c$ by fitting the straight line near the contact as shown in Fig.~\ref{fig:schematic_vesicle_adhesion}B. We did a similar angular averaging to calculate the contact angle.

\begin{figure}
    \centering
    \includegraphics[width=0.995\linewidth]{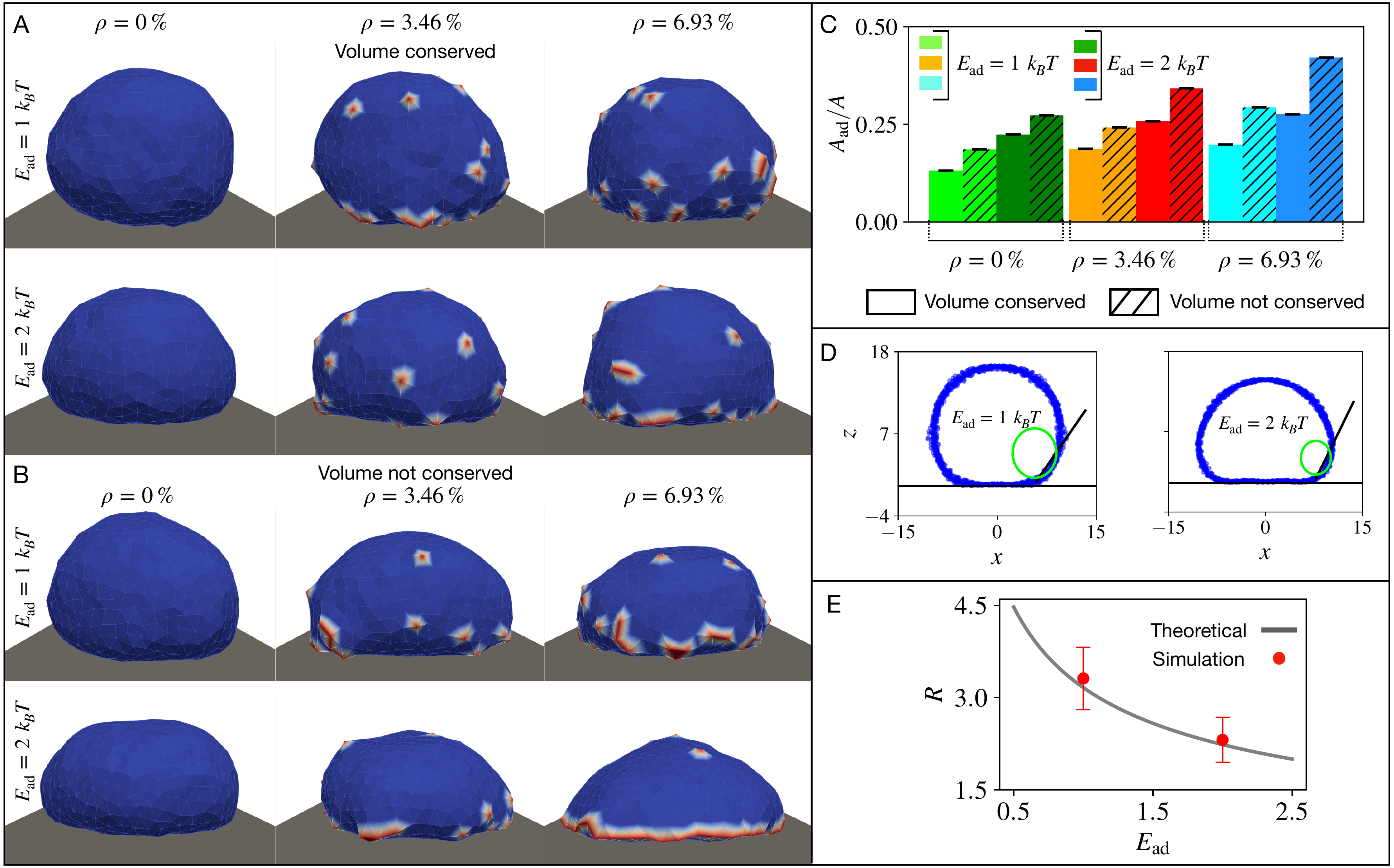}
    \caption{Verification of the radius of curvature near the contact line of a vesicle on the flat substrate. A) Snapshots of the vesicle adhering to the flat substrate when the volume and the area are conserved, for different concentrations of CMC (passive) and adhesion energies. B) Same as (A)  when the volume conservation constraint is removed. C) Adhered area fraction is shown as a bar plot for different cases shown in (A,B). D) Numerical extraction of the radius of curvature near the substrate-vesicle contact line from the simulated shapes. E) Verification of the predicted dependence of the radius of curvature (as defined in Fig.\ref{fig:schematic_vesicle_adhesion}) on the adhesion energy ~\cite{marevs2012determination} for two cases of $E_{\rm ad}=1k_BT,~2k_BT$, when there are no CMC ($\rho=0\%$) and the volume is conserved.}
    \label{fig:single_sphere_verify}
\end{figure}

A single vesicle is allowed to spread on an adhesive substrate for different adhesion energy $E_{\rm ad}$ parameters and different CMC densities $\rho=0,~3.46\%,~6.93\%$. The shapes for all these different cases when the total volume of the vesicle is conserved are shown in Fig.~\ref{fig:single_sphere_verify}A. We showed the case when the vesicle volume is not conserved in Fig.~\ref{fig:single_sphere_verify}B. We have shown the steady-state adhered area fraction for all these cases in Fig.~\ref{fig:single_sphere_verify}C. There is a higher adhered area fraction when the adhesive energy $E_{\rm ad}$ per node is higher. We also see that the curved-protein percentage is  driving the spreading, by reducing the bending energy cost near the contact line. 
When the volume is not conserved, we find that the vesicle can lose its volume to assume a sheet-like shape and spread more, as shown in Fig.~\ref{fig:single_sphere_verify}C. 

We have shown the two-dimensional side cut of the vesicle on $x-z$ plane and the fitted circle in order to calculate the radius of curvature in lime (See Fig.~\ref{fig:single_sphere_verify}D).
We verified the radius of curvature $R$ for $E_{\rm ad}=1k_BT$ and $2k_BT$ when volume is conserved and the curved proteins are absent as shown in Fig.~\ref{fig:single_sphere_verify}E. The theoretical estimation is given by the grey solid line .

\section{Verification of radius of curvature near the contact with two identical vesicles}
\begin{figure}
    \centering
    \includegraphics[width=0.75\linewidth]{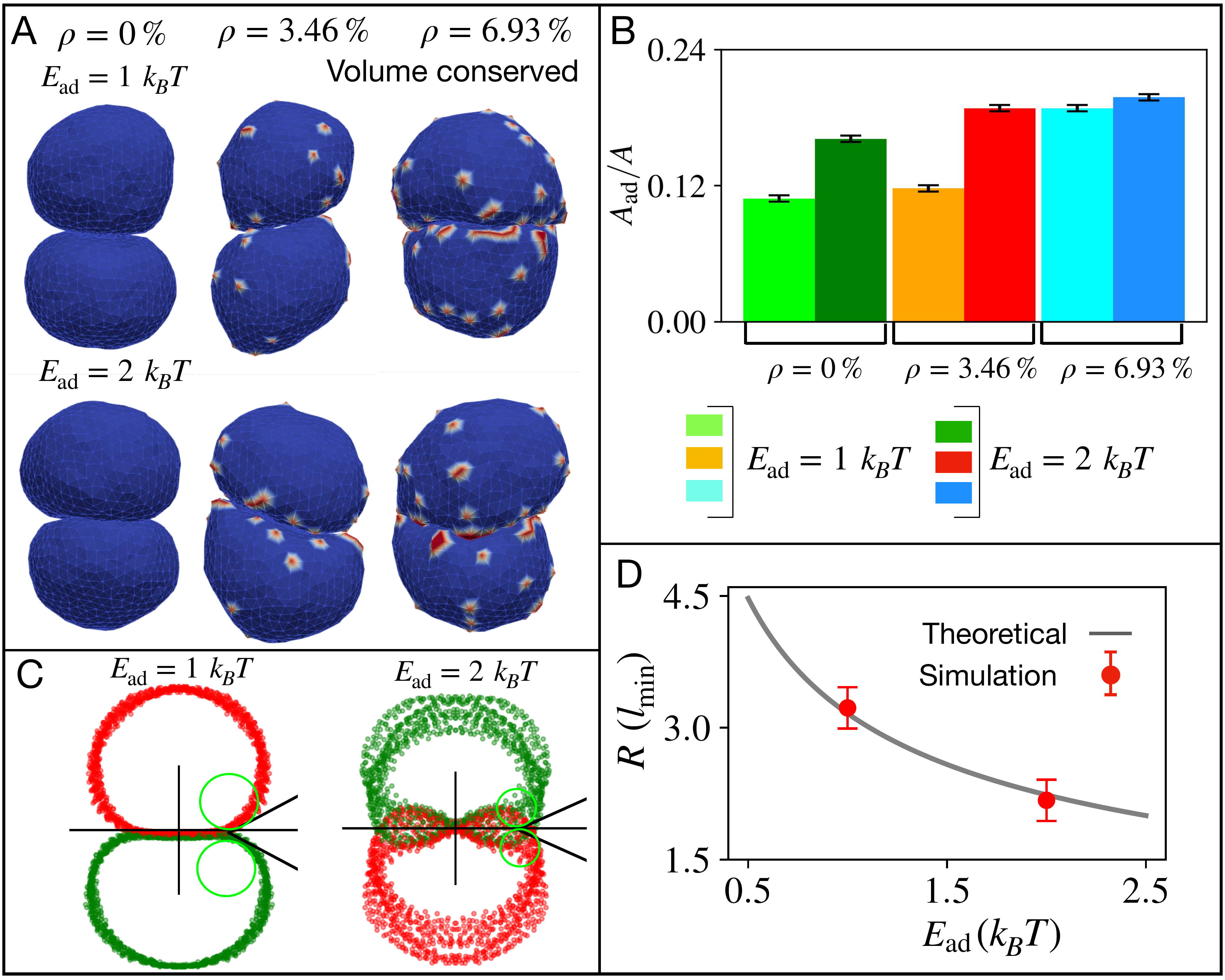}
\caption{Verification of radius of curvature near the contact line of two adhered symmetric vesicles when the volume and area are kept constant. A) Snapshots of two vesicles adhering to each other for three different CMC densities ($\rho=$ 0\%, 3.46\%, and 6.93\%), and different adhesion energies $E_{\rm ad}=1k_BT,~2k_BT$. B) The adhered area fraction is shown as a bar plot. C) The numerical calculation of the radius of curvature near the contact line between the two adhered vesicles  for the case of $\rho =0\%$. D) Comparison of the numerical estimates of the radius of curvature compared with the theoretical prediction $R_{\rm th}=\sqrt{\frac{\kappa}{2 E{\rm ad}}}$ shown in grey line. We used the number of vertices $N=722,~\kappa=20 k_BT,~w=1k_BT,~c_0=1l_{\rm min}^{-1}$ for both vesicles. }
    \label{fig:double_sphere_verify}
\end{figure}

Next, we consider two identical vesicles that adhere to each other. Again, we find the radius of curvature near the contact and validated with the previous work ~\cite{marevs2012determination}.
Though the technique to calculate the radius of curvature is similar to the case of the single vesicle, you need to orient the vesicles properly before making any observation. We first draw a straight line joining the center of mass of the two vesicles, as shown in Fig.~\ref{fig:schematic_vesicle_adhesion}C.
We then align this line to the $z$ axis by rotation. We calculated the radius of curvature near the contact for the cases of no CMC and with volume conservation, for two different parameters for the adhesion energy $E_{\rm ad}$ per node. The snapshots for two vesicles spreading on each other are shown in Fig.~\ref{fig:double_sphere_verify}A. The adhesive area fraction is increased as the adhesive energy per node is increased (See Fig.~\ref{fig:double_sphere_verify}B). Also, the adhesion area fraction is larger in the presence of the CMC. The volume conservation hinders the spreading, with the adhered area fraction increasing with increasing CMC concentration and increasing adhesive energy $E_{\rm ad}$ per node. 
Fig.~\ref{fig:double_sphere_verify}C shows the side cut of the two vesicles adhering to each other for the no CMC and volume conservation case. The verification of the radius of curvature near the contact point is shown in Fig.~\ref{fig:double_sphere_verify}D.
\section{Adhesion dynamics between two vesicles with passive CMC and no volume conservation}

\begin{figure}
    \centering
    \includegraphics[width=\linewidth]{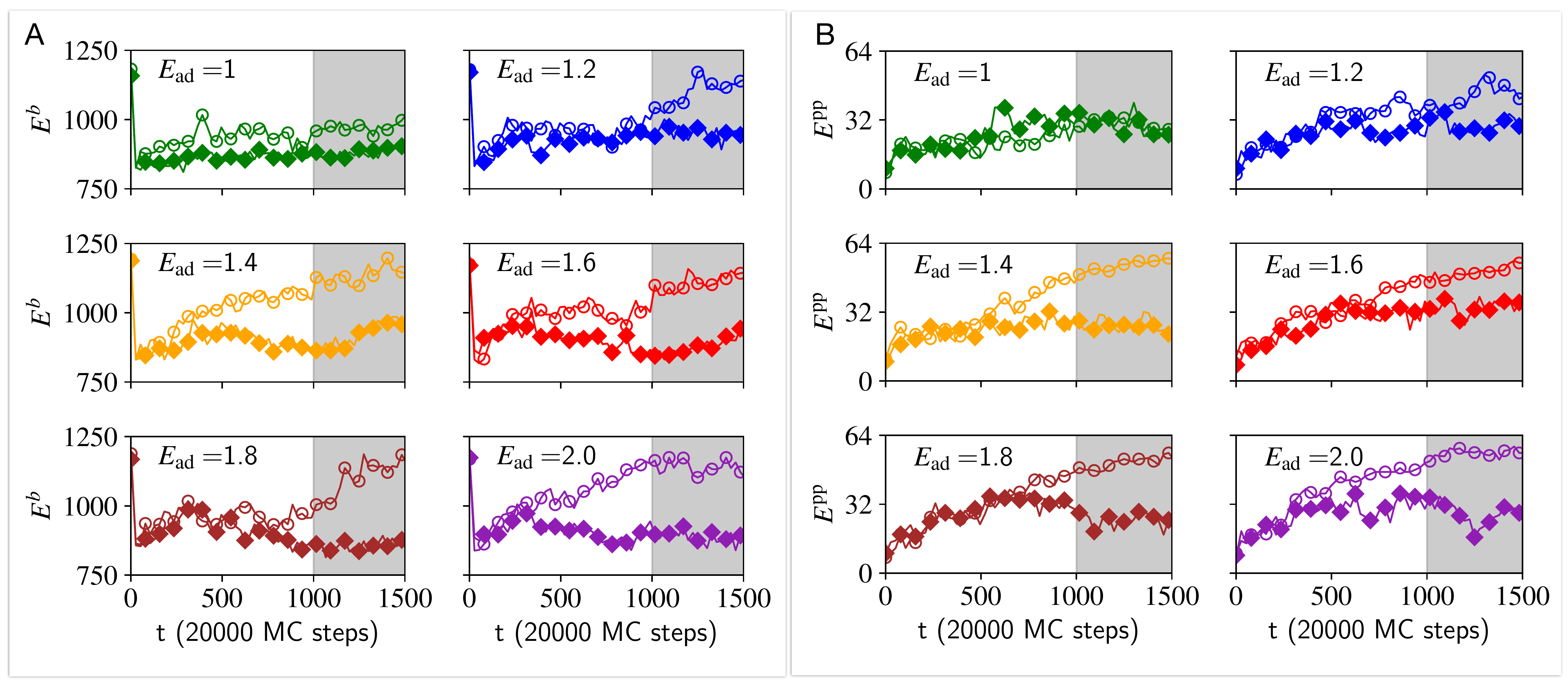}
    \caption{Energy terms during the adhesion dynamics between two vesicles shown in Fig.2 (open symbols for the top vesicle $2$, and filled symbols for the bottom vesicle $1$). A) The asymmetry in the bending energies $E^b$ for two identical vesicles without volume conservation as the adhesive energy strength increases from $1k_BT$ to $2k_BT$. The bending rigidity for two vesicles is set to $\kappa=20 k_BT$ and the protein percentage is set to $\rho=6.92\%$ for both vesicles. B) The asymmetry in protein-protein interaction energy $E^{\rm pp}$ for the same cases as in (A). Each vesicle consists of 722 vertices.}
    \label{fig:symmetry_break_extra_energy}
\end{figure}

In Fig.2 we show the steady-state final shapes of two identical vesicles adhered to each other for different strengths of
adhesion energy parameter $E_{ad}$ and passive CMC. In Fig.\ref{fig:symmetry_break_extra_energy} we plot the time evolution of the energy terms for the two vesicles during these processes. We show that above a critical adhesion energy of $\sim1.2 k_BT$ there is spontaneous breaking of the symmetry, with one of the vesicles experiencing a large increase in its bending energy (Fig.\ref{fig:symmetry_break_extra_energy}A), which is of course compensated by the large increase in the negative magnitude of the adhesion energy (Fig.2B,C). The vesicle that deforms into a cap-shape forms a ring cluster of its CMC, which indeed provides a larger interaction energy between the proteins (Fig.\ref{fig:symmetry_break_extra_energy}B).

\section{Benchmark of phagocytosis with a rigid spherical particle}
Next, we have two non-identical vesicles and let them interact through adhesion, and the active forces that mimic the actin cytoskeleton. One vesicle is bigger in size and contains CMC, representing a cell-like vesicle. The other vesicle is smaller in size and has no CMC for simplicity, serving as the target-vesicle. We kept the bending rigidity of the cell-like vesicle $\kappa=20k_BT$ throughout the paper. Therefore, whenever we talk about the bending rigidity $\kappa$ in the context of the phagocytosis process, it is the bending rigidity of the target vesicle. 

Before we explore the complex behaviours exhibited by the very soft objects due to their big deformations, we validate the model by comparing two cases: We consider a completely rigid spherical object as a target \cite{sadhu2023theoretical}, and compare the dynamics to the case of a very rigid target (high bending rigidity $\kappa=2000 k_BT$) vesicle of the same radius. We simulate these two cases with a model cell-like vesicle containing wither passive ($F=0k_BT ~l_{\rm min}^{-1}$) or active ($F=2 k_BT~l_{\rm min}^{-1}$) CMC.

\begin{figure}
    \centering
    \includegraphics[width=\linewidth]{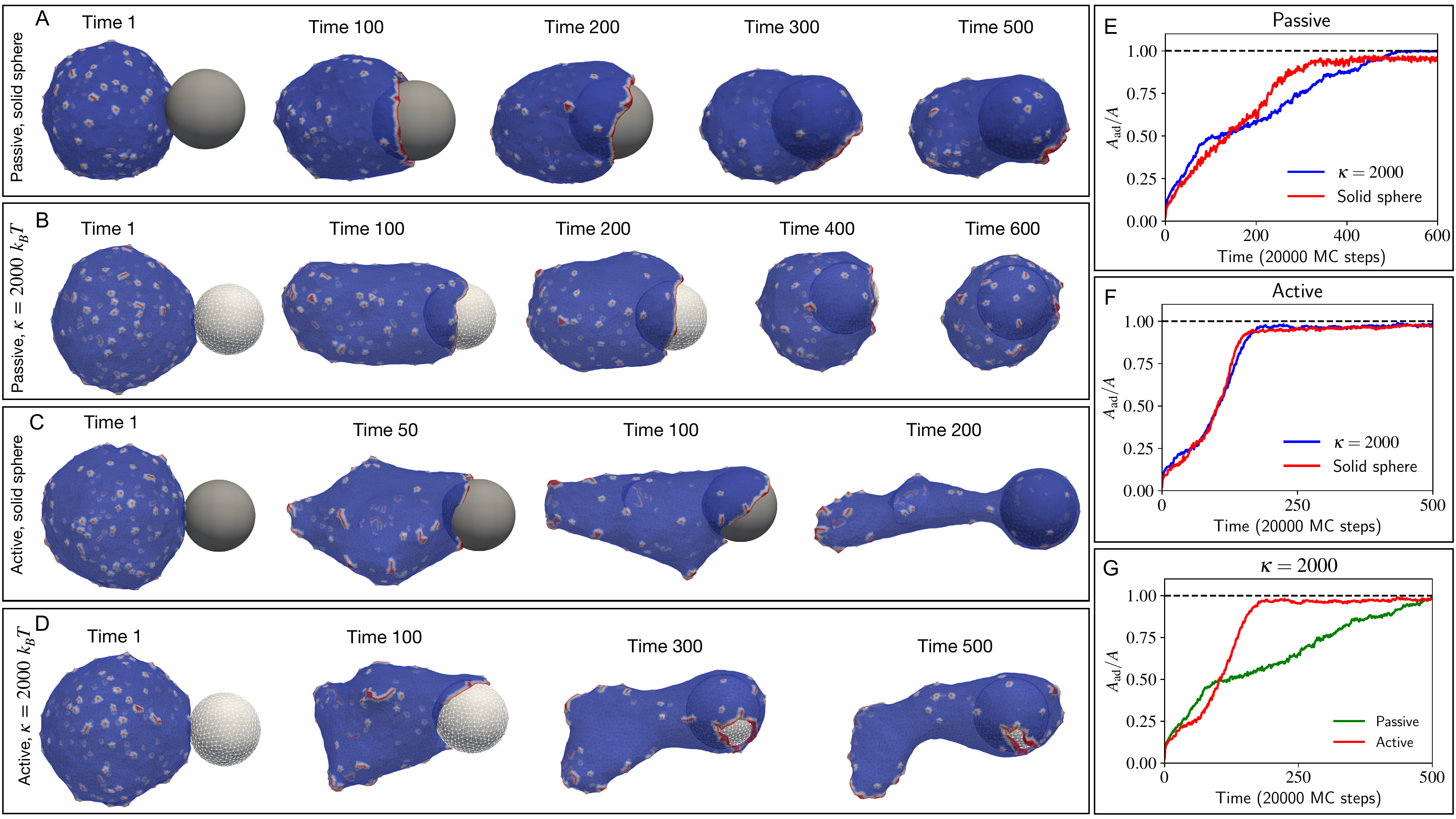}
\caption{The snapshots of the phagocytosis process when the big vesicle with passive proteins (no. of vertices $N=3127$, no. of passive proteins $N_p=150$, and bending rigidity $\kappa=20~k_BT$) tries to engulf a (A) rigid sphere of radius $R=10~l_{\rm min}$ and (B) a vesicle (nearly a sphere of radius of $R\approx 10~l_{\rm min}$ in shape) with high bending rigidity $\kappa^T=2000k_BT$. Next, the big vesicle is active, it can exert force on the target. The active force parameter $F=2~k_BT~l^{-2}_{\rm min}$. The snapshots are shown when the target is (C) a rigid sphere of radius $R=10~l_{\rm min}$ and (D) a vesicle with high bending rigidity $\kappa^T=2000k_BT$. Comparison of the adhesive area fraction of the target between the rigid sphere and the vesicle of bending rigidity $\kappa^T=2000 ~k_BT$ when the engulfing vesicle is (E) passive $F=0$ and (F) active case $F=2~k_BT~l^{-2}_{\rm min}$. (G) Comparison of the adhesive area fraction of the target between the passive and active engulfment.}
    \label{fig:adhesion_compare}
\end{figure}

We find the adhered area fraction is very similar for the completely rigid object and the target vesicle with very high bending rigidity $\kappa=2000k_BT$ for both the passive and active cases, respectively, as shown in Fig.~\ref{fig:adhesion_compare}E-F.
We have also found that the active vesicle is more efficient in complete engulfment compared to the passive vesicle, as shown in Fig.~\ref{fig:adhesion_compare}G. We have shown the different stages of the engulfment processes with snapshots of the interaction between two vesicles in Fig.~\ref{fig:adhesion_compare}A-D for four different cases.

\section{Vesicle area deformation and fluctuations during passive engulfment}

In Fig.3 we show the engulfment of vesicles of different bending rigidities by cell-like vesicle that contains passive CMC. The engulfment involves deformations of the target vesicle, which are quantified as follows: First, we find the centroid $\overrightarrow{r}_0$ of the vesicle by averaging the position vectors of all the vertices. Next, we find the distances of all the vertices from the centroid that is given by,
\begin{equation}
    \tilde{r_i}=|\overrightarrow{r}_i-\overrightarrow{r}_0|.
    \label{eq:distance_from centroid}
\end{equation}
Next, we find the relative standard deviation of these distances $r_i$ is given by, 
\begin{equation}
    \Delta R/R=\sqrt{\langle \tilde{r_i}^2\rangle -\langle \tilde{r_i}\rangle^2}/\langle \tilde{r_i}\rangle.
    \label{eq:roundedness}
\end{equation}
A sphere has a non-roundedness measure of zero, and it increases as the vesicle becomes more non-spherical. 

\begin{figure}
    \centering
    \includegraphics[width=\linewidth]{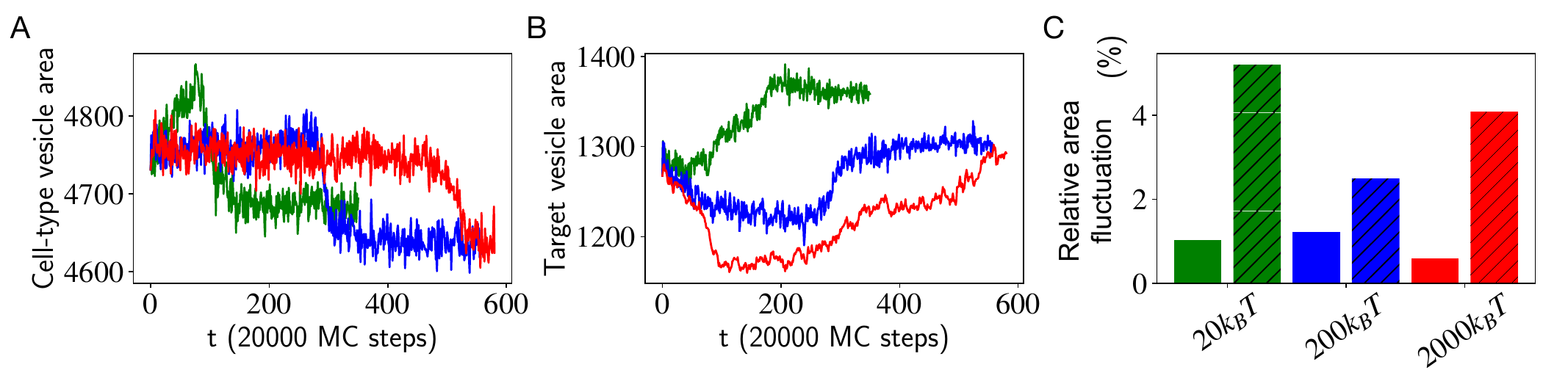}
    \caption{The evolution of the vesicles' area during the engulfment simulations for the passive case (Fig.3): A) The cell-like vesicle area over time, for three different bending rigidity values of the target vesicle: $\kappa=20,~200,~2000$ respectively in the units $k_BT$ (green, blue and red respectively). B) The target vesicle's area over time, as in (A). C) The relative fluctuation of the area for both vesicles is calculated with respect to the initial area of the vesicle. The smooth-color bars are for the cell-like vesicle, while the diagonal hatched bars are for the target vesicle.
    }
    \label{fig:area_fluctuation}
\end{figure}

The non-roundness parameter for the target vesicle is shown in Fig.3C. Similarly, changes to the vesicle's area during this engulfment process are shown in Fig.\ref{fig:area_fluctuation}.

\section{Detachment of the cell-like vesicle from the target after a pushing event}
\begin{figure}
    \centering
    \includegraphics[width=0.5\linewidth]{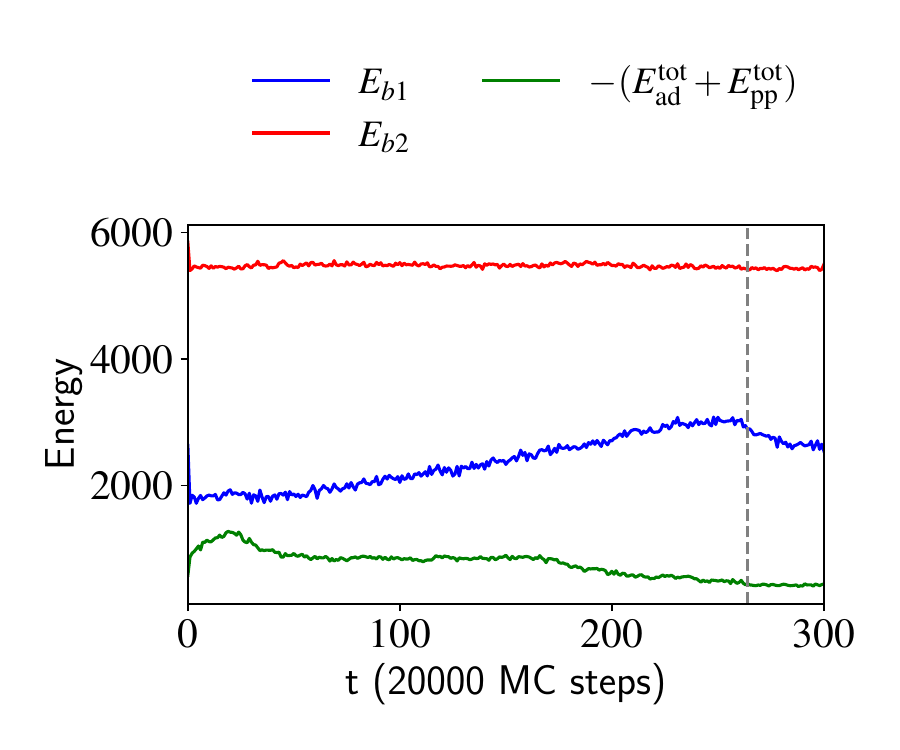}
    \caption{The timeseries for the energy terms for the case of pushing (Fig.4D), when the bending rigidity of the target is $\kappa=200~k_BT$. The bending energy of the target vesicle is nearly constant over time (red). We show the competition between the bending energy of the cell-like vesicle (blue) and its adhesion and protein-protein interaction energies (green). The grey dashed line indicates the time when the cell-like vesicle detached from the target vesicle. This detachment allows the cell-like vesicle to retract its long thin protrusion, thereby reducing its bending energy.}
    \label{fig:energy_pushing}
\end{figure}

In Fig.~4D we showed that for intermediate target vesicle rigidities the cell-like vesicle ends up pushing it. This process ends with the cell-like vesicle detaching from the target vesicle and retracting the long protrusion that formed. In Fig.\ref{fig:energy_pushing} we plot the bending energies of the two vesicles during this process, and the sum of the adhesion and protein-protein interaction energy for the cell-like vesicle. We see that as the pushing event progresses, there is an increase in the bending energy associated with the long protrusion of the cell-like vesicle ($E_{b1}$ in Fig.\ref{fig:energy_pushing}), while the adhesion and protein-protein binding energies which compensate for it decrease. This increasing energy cost eventually drives the detachment, and retraction, which decreases the bending energy cost.

\section{Effect of holding the target on the engulfment process}
We have found that in the regime of intermediate rigidity, the cell-like vesicle pushes the target vesicle and ultimately leaves it without biting or engulfing it (Fig.~6E, Fig.~4D). Here we simulated the dynamics when an external constraint confines the target vesicle. For example, in an ``in vitro" experiment a target may be held by a pipette \cite{cornell2025target}, or cells may be confined by surrounding cells inside tissues. 

To mimic such a confinement effect, we designated a patch of vertices on the target vesicle that are not allowed to move. We choose a patch of vertices centered at the pole of the target vesicle away from the cell-like vesicle (black patch on the right-most edge of the target vesicle in Fig.~6F). 
This is a patch of depth of $l_{\rm min}$ from the pole of the target vesicle. We found that this confinement can change the dynamics from pushing to engulfment or biting (trogocytosis), as shown in Fig~6 (E-F) and see the Movie S11. 

We showed how different quantities, such as the adhered area fraction, deformation, tangential force fraction and normal force fraction evolve over MC steps as shown in Fig.~6 (G-I). Holding a patch of the target vesicle changes the entire engulfment dynamics by directing more of the active force to be tangential to the target vesicle, therefore much more efficient in driving engulfment.

\section{Effect of internal pressure of the target cell}
When an osmotic pressure difference $p$ exists between the inside and the outside of the target vesicle, an additional energy cost term is included \cite{Fonari2019}
\begin{equation}
    \Delta E_{\rm p}=p dV
\end{equation}
Here, $dV$ is the change in volume due to the Monte Carlo moves (e.g. vertex movement, bond flip).
The positive internal pressure $p$ creates more tension on the membrane, and the vesicle gets inflated.

\begin{figure*}
    \centering
    \includegraphics[width=\linewidth]{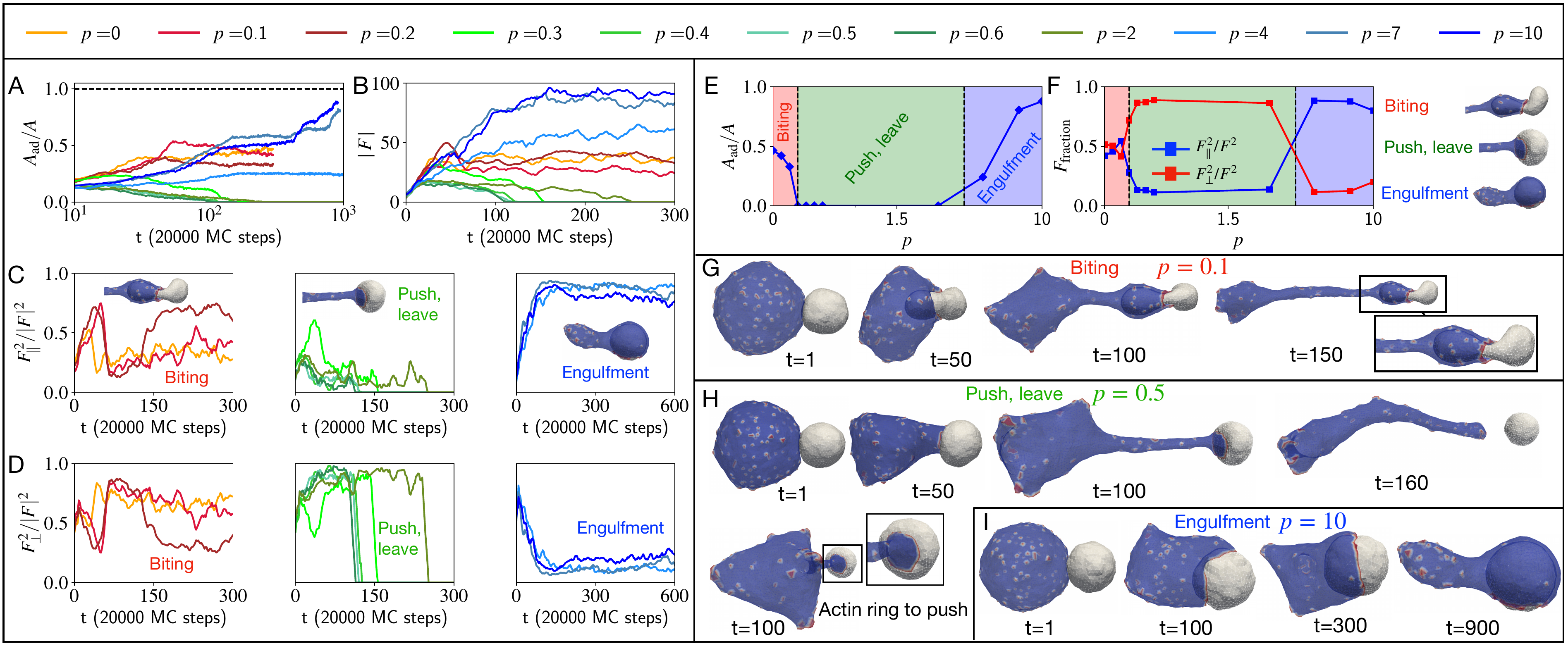}
    \caption{Effect of the internal pressure of the target cell on the process of phagocytosis. The smaller vesicle (target) and the bigger vesicle (attacking cell) are made of $847$ and $3127$ vertices, respectively. The time evolution of the adhesive area fraction and magnitude of the force applied on the target cell by the attacking cell are shown in A) and B), respectively. C) It shows the fraction of tangential force on the target for three different cases of nibbling, push-leave, and engulfment in three different panels. D) Similarly, the fraction of normal or pushing force on the target for three different cases of nibbling, push-leave, and engulfment in three different panels. E) We have shown the steady state average value of adhesive area fraction with the bending rigidity $\kappa$ of the target cell. F) Here, we averaged the tangential and normal force fraction in the relevant time windows. (G-I) The time evolution of the shapes and the snapshots are shown for three different cases of nibbling, push-leave, and engulfment respectively.}
    \label{fig:transition_via_pressure}
\end{figure*}

We have systematically studied the effect of the internal osmotic pressure of the target vesicle on the engulfment process, while keeping the bending rigidity of the target-like vesicle at $\kappa=20k_BT$. We found three dynamical phases (Fig.~\ref{fig:transition_via_pressure}), identical to the phases found for varying the bending rigidity (Fig.4).
As the internal pressure increases, the target vesicle membrane is under higher tension, which opposes the deformation and keeps it more spherical. Increasing internal osmotic pressure in the target vesicle therefore affects the engulfment process in an identical way to increasing the bending rigidity of the target vesicle (Fig.4).

\begin{figure}
    \centering
    \includegraphics[width=0.95\linewidth]{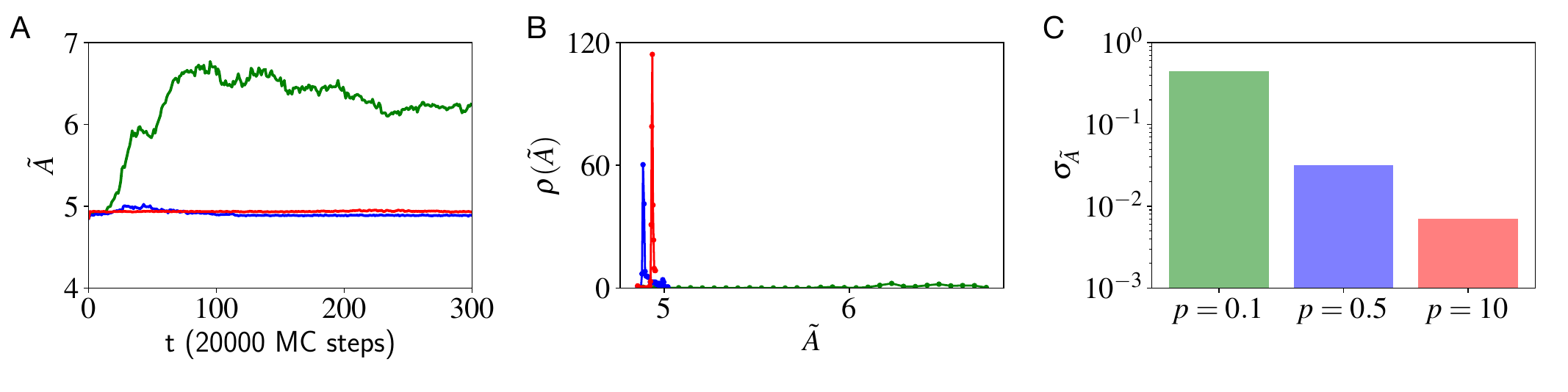}
\caption{A measure of membrane tension when the area and the volume both change over time during the active engulfment of a target-like vesicle with different internal pressure (Fig7G-I). A) The size normalised area $\tilde{A}$ (Eq.\ref{tildeA}) for three different pressure differences $p=0.1,~p=0.5,$ and $p=10$ shown in green, blue and red, respectively. B) The probability distribution of the size normalized area  $\rho(\tilde{A})$. C) The standard deviation $\sigma_{\tilde{A}}$ of the size-normalized area. Higher tension results in lower standard deviation of the area fluctuations, and therefore higher internal pressure implies higher membrane tension. We set the bending rigidity of both the vesicles to $20k_BT$. The active force parameter $F=2k_BT$, adhesion energy per node is $E_{\rm ad}=2k_BT$ between the two vesicles.}
    \label{fig:tension_measure}
\end{figure}

We next calculated a measure of the membrane tension due to the internal osmotic pressure $p$, as shown in Fig.\ref{fig:tension_measure}. As both the volume and area of the vesicle are changing, we introduced the volume-normalized area $\tilde{A}$ given by,
\begin{equation}
    \tilde{A}= \frac{A}{V^{2/3}}
    \label{tildeA}
\end{equation}
where, $A$ and $V$ are the area and the volume of the vesicle. The time evolution of $\tilde{A}$ is shown in Fig.~\ref{fig:tension_measure}A for the cases of internal pressures of $p=0.1,~0.5$ and $10$, during the interaction of the target with an active cell-like vesicle (as in Fig.7G-I). Next, we calculated the probability distribution of the volume-normalized area $\rho(\tilde{A})$ (Fig.\ref{fig:tension_measure}B). As pressure increases, the distribution function has a very narrow and sharp peak. A sharper peak implies a lower standard deviation $\sigma_{\tilde{A}}$ of the area fluctuations for higher internal pressure $p$ as shown in Fig.~\ref{fig:tension_measure}C. Lower area fluctuations are equivalent to higher membrane tension.


\begin{figure}
    \centering
    \includegraphics[width=\linewidth]{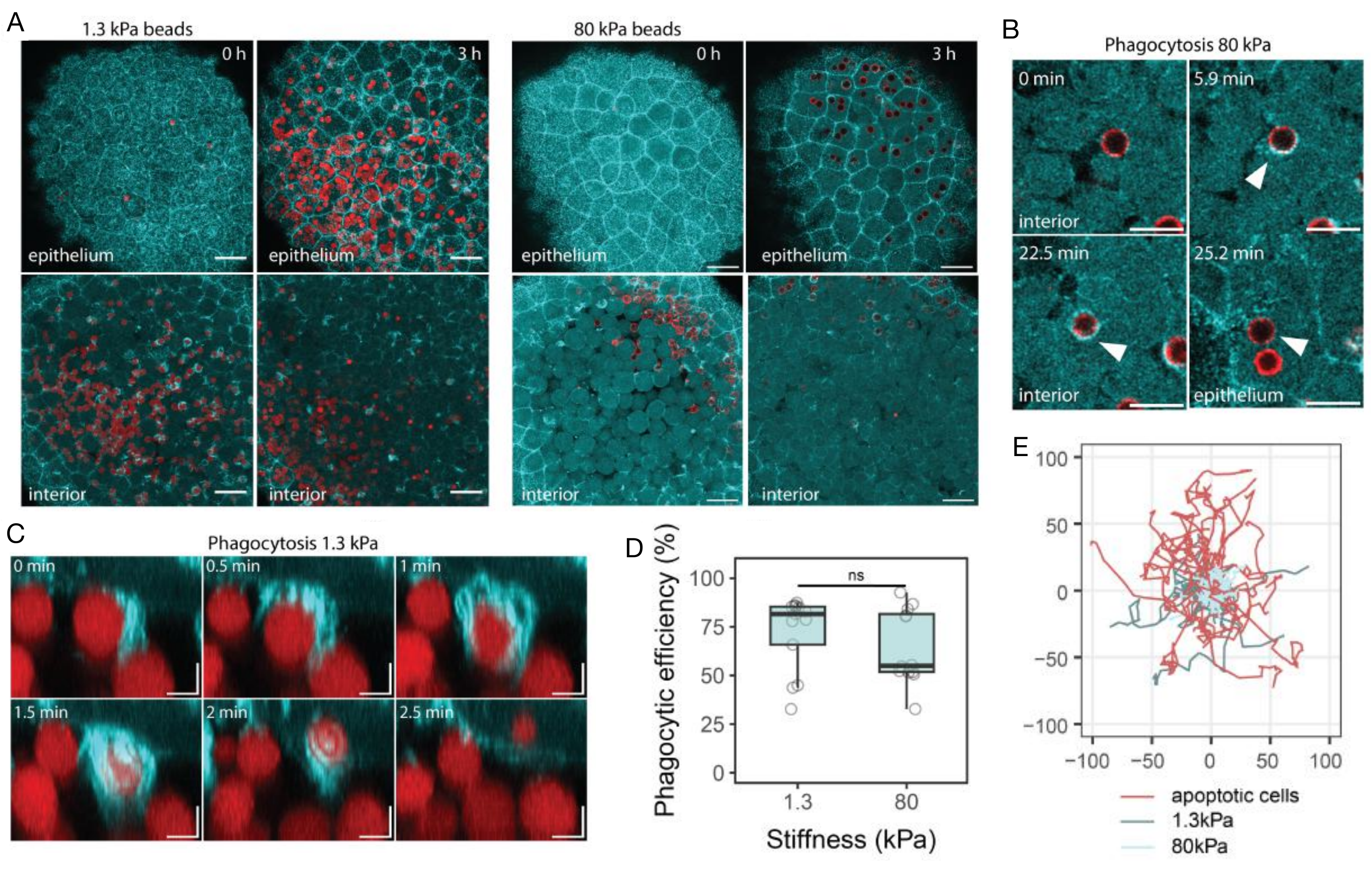}
    \caption{\textbf{Epithelial phagocytosis and mobility of apoptotic cells and synthetic targets with variable target stiffness in vivo} --- A) Representative images showing the presence of synthetic apoptotic targets (1.3 kPa, left; 80 kPa, right) in the interior of a zebrafish blastula stage embryo expressing Lifeact-GFP (cyan) at t=0 and their phagocytic clearance by the outer epithelial tissue over time (t=3 h). B) Representative images of an individual phagocytosis event of 80 kPa hydrogel beads in vivo. C) High-resolution transversal view of an individual phagocytic uptake event of 1.3 kPa hydrogel beads in a Lifeact-GFP (cyan) expressing embryo. D) Quantification of phagocytic efficiency derived as the percentage of synthetic targets cleared by the epithelium for 1.3 kPa hydrogel beads (n = 13 embryos) and 80 kPa hydrogel beads (n = 12 embryos). Data points represent the apoptotic target clearance efficiency as the percentage of cleared targets in individual embryos. N = 3 independent experiments. Welch Two Sample t-test, p = 0.99947. E. Tracks of individual apoptotic targets showing the path travelled over a maximum period of 30 min for apoptotic cells (red, n = 32 tracks from 3 embryos), 1.3 kPa hydrogel beads (dark cyan, n = 29 tracks from 3 embryos) and 80 kPa hydrogel beads (light cyan, n = 30 tracks from 3 embryos). Tracks were centered to the origin. Scalebars: 40$\mu$m (A), 20$\mu$m (B), 5$\mu$m z: 5$\mu$m (C).}
    \label{fig:placeholder}
\end{figure}

\section{Implementation of bulk modulus}\label{sec:bulk}

\begin{figure*}
    \centering
    \includegraphics[width=\linewidth]{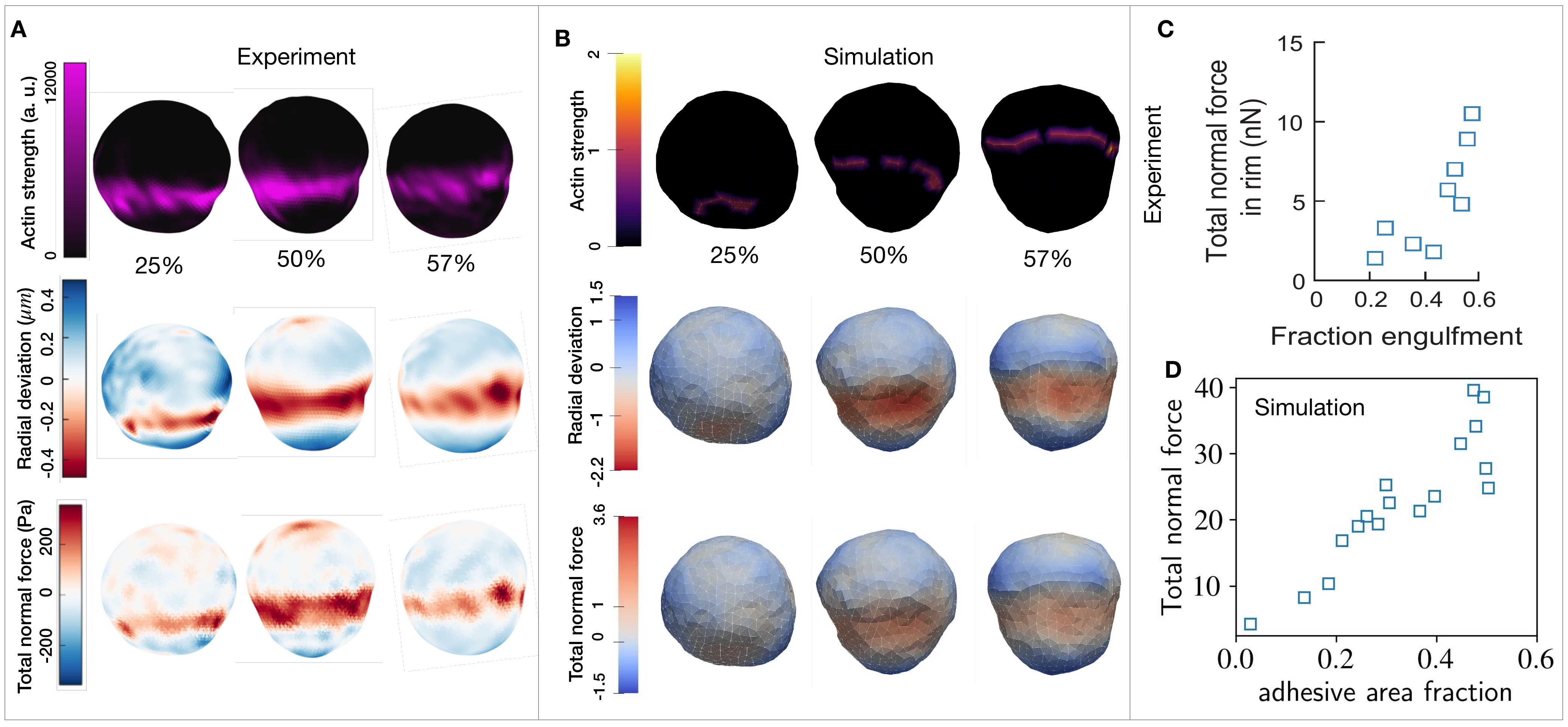}
    \caption{A) 3D reconstructions of deformable acrylamide-co-acrylic acid-microparticles (DAAM-particles) (1.4 kPa, 9 µm) revealing F-actin over the particle surface, detailed target deformations induced during phagocytosis and normal forces inferred from shape deformations. DAAM-particles were functionalized with Immunoglobulin G (IgG) and stained with TAMRA-cadaverine and exposed to phagocytosis by RAW264.7 cells, which were then fixed and stained for F-actin.  Each panel displays (left) F-actin distribution across the particle surface, (middle) radial deformation, and (right) the normal traction forces inferred from the shape deformations. (B) Snapshots of the simulated target vesicle at different values of engulfed fraction. Heatmaps show the location of the CMC (left), target vesicle radial deformation (middle) and the normal component of the active force exerted by the CMC on the target vesicle (right). These correspond to the actin localization in the experiments, bead deformation and normal elastic forces, respectively. In these simulations, we added the energy due to the bulk modulus of the target $\kappa_{bulk}=0.5~k_BT/l^2_{\rm min}$ (Eq.\ref{fbulk}), while the bending modulus of the target is $\kappa=20~k_BT$. (C,D) plots of total normal force as function of the adhesive area fraction, in the experiment and simulation, respectively. }
    \label{fig:daan_combined}
\end{figure*}

In order to compare with the experiments of phagocytosis of elastic beads \cite{vorselen2020microparticle,settle2024beta2} at a high resolution (Fig.~\ref{fig:daan_combined}), we need to implement a description of the bulk elasticity of the target vesicle.

We implemented a bulk modulus for the target cell in a simple way. The desired shape for the vesicle is a sphere with a radius $r_0=10~l_{\rm min}$. All the vertices feel a force or cost some energy if it deviates from that shape. The force is given by,
\begin{equation}
    \bold{F}_i^{\rm bulk}=\kappa_{\rm bulk}(r_i-r_0) \hat{n}_{\rm bulk}
    \label{fbulk}
\end{equation}

where, $r_i$ is the distance of the center of mass from the $i$th vertex, and $\hat{n}_{\rm bulk}$ is the unit vector directed to the center of mass from the $i$th vertex.  We added the energy due to the bulk modulus of the target that is set to $\kappa_{bulk}=0.5~k_BT/l^2_{\rm min}$. We simulated the engulfment of soft target of $\kappa=20k_BT$ by a cell-like vesicle and measured the actin strength, radial deviation, and the normal force along the formed phagocytic cup or actin ring. We compared these quantities with the experimental data which is done with artificial DAAM particles, as shown in Fig.~\ref{fig:daan_combined} (A-B).
We compared the total normal force with the fraction of the adhesion area of the target or the fraction of engulfment.
The overall agreement is very good. Note that the actin ring in the experiments is necessarily wider than the ring of CMC in the model, since the CMC denote only the leading edge nucleators of actin and not the whole actin network that forms behind them.

In the experiments at the bead's pole pointing at the engulfing cell a pulling force acting on the bead is measured (Fig.~\ref{fig:daan_combined}A), which may arise from contractile forces pulling on these adhesion sites, or due to the effect of actin treadmilling emanating from the leading edge, exerting forces that pinch the bead at this pole (similar to actin treadmilling-induced forces that are involved with endocytosis  \cite{motahari2017actin}). This effect is absent from our simple model. 

\section{Details of Experimental and simulation methods}

\textbf{Lipids}---Phosphocholine (PC) lipids (Avanti Polar Lipids) were used as purchased without further purification. Lipid stock solutions in chloroform contained a ternary mixture of 98 mol$\%$ POPC, 1 mol$\%$ biotin-PE, and 1 mol$\%$ PEG2K DSPE. GUVs are diluted in an ionic solution of PBS and all lipids in our mixtures are zwitterionic. We added PEG2K DSPE to block GUVs from aggregating in the charge-screened PBS solution. \\

\textbf{Antibodies}---Antibodies used to opsonize GUVs were purchased from Santa Cruz Biotechnology and used without further labeling or purification. Biotin was bound by AlexaFluor647-labeled anti-biotin mouse IgG (clone BK-1/39, Santa Cruz Biotechnologies). \\

\textbf{RAW 264.7 cell culture}---RAW 264.7 murine male macrophage-like cell line was obtained from and authenticated by the UC Berkeley Cell Culture Facility. Cells were cultured in RPMI 1640 media (Corning) supplemented with $10~\%$ heat-inactivated fetal bovine serum (HI-FBS, Thermo Fisher Scientific) and $1~\%$ Pen-Strep (Thermo Fisher Scientific). RAWs were cultured in non-tissue culture-treated 10 cm dishes (VWR) at 37$^o$C, 5$~\%$ $CO_2$. \\

\textbf{Stable LifeAct GFP RAW 264.7 cell line}---HEK293T cells were grown in a 6-well plate to 80$\%$ confluency, and 160 ng VSV-G, 1.3 $\mu$g CMV 8.91, and 1.5 $\mu$g target pHR LifeAct GFP expression vector were transfected into HEK293T cells using TransIT-293T transfection reagent (Mirus Bio). Viral supernatants were collected 60 hours after transfection and spun at 4000 G to remove HEK293T cells. Viral supernatant was stored at 4$^o$C for no longer than 48 hours prior to infection. For lentiviral infection, 500 $\mu$L of viral supernatant was added to 5e5 RAW 264.7 macrophages along with 4 $\mu$g/mL polybrene, and cells were spun at 400G for 25 minutes at 37$^o$C and then resuspended and plated in a 6-well plate. Viral media was replaced with fresh growth media 24 h after infection. Cells were sorted via fluorescence-activated cell sorting on an Influx Cell Sorter (Beckton-Dickinson), and a population of cells expressing LifeAct GFP was expanded and frozen for later use. \\

\textbf{Synthesis of deformable acrylamide acrylic acid particles}---Hydrogel particles were synthesized as previously described. Briefly, acrylamide mixtures of acrylamide (AAm), acrylic acid (AAc), crosslinker $N,N'$-methylenebisacrylamide (BIS), 150 mM NaOH, 0.3\% (v/v) tetramethylethylenediamine (TEMED), 150 mM MOPS (prepared from MOPS sodium salt, pH 7.4) were prepared. Total mass concentration of acrylic components ($C_{\rm AAm}+C_{\rm AAc}+C_{\rm BIS}$) was 100 mg/mL respectively 200 mg/mL and crosslinker concentration ($C_c=m_{\rm BIS}/(m_{\rm AAm}+m_{\rm AAc}+m_{\rm BIS})$) 0.64\% respectively 5.06\%, for 1.3 kPa and $\sim$80 kPa particles. Prior to extrusion, the mixture was degassed for 15 min and kept under nitrogen atmosphere. Tubular hydrophobic Shirasu porous glass (SPG) membranes of 20 mm length with pore size (diameter) 1.1 $\mu$m respectively 2.4 $\mu$m were used for soft and stiff particles. Membranes were sonicated under vacuum in HPLC grade n-heptane to remove gas trapped in the membrane. The membranes were mounted on an internal pressure micro kit extruder (SPG Technology) and immersed into an oil phase ($\sim$125 mL) consisting of hexanes (99\% ACS reagent, mixed isomers) and 3\% (v/v) Span 80 (Sigma Aldrich, S6760). 10 mL of gel mixture was extruded through SPG membranes under nitrogen pressure of $\sim$30 kPa, while the oil phase was continuously stirred and kept under nitrogen atmosphere in a 3-neck water-jacketed flask. After completion of extrusion, the emulsion temperature was increased to 60$^\circ$C. Once the temperature equilibrated, DAAM-particle polymerization was induced by addition of $\sim$225 mg 2,2'-Azobisisobutyronitrile (AIBN) (1.5 mg/mL final concentration). The polymerization reaction was continued for 3 h at 60$^\circ$C and then at 40$^\circ$C overnight. Polymerized particles were subsequently washed (5× in hexanes, 1× in ethanol), dried under nitrogen flow for $\sim$30 min, and resuspended in PBS, pH 7.4. Soft and stiff particles had diameters of 8.9 respectively 8.2 mm, as determined by phase-contrast imaging and image analysis in ImageJ.\\

\textbf{Microparticle traction force microscopy (MP-TFM) analysis of phagocytosis of  DAAM-particles}---
Experimental data was obtained from prior conducted experiments~\cite{Daan_2021_figshare} in which RAW 264.7 macrophages were exposed to IgG-functionalized 1.3 kPa deformable poly-Aam-co-AAc microparticles (DAAM-particles), stained for F-actin using Alexa Fluor-488 conjugated phalloidin, and imaged by confocal microscopy~\cite{Vorselen2021_elife}. DAAM-particle 3D shape reconstructions and force analysis were performed as previously described \cite{Vorselen2020_nat_comm, Mali2025_nat_protocols}. Briefly, the inverse problem of inferring the traction forces \textbf{\textit{T}} is solved iteratively until a minimal gradient tolerance is reached. During this optimization, an ideal sphere with the same particle volume of the measured individual particle is subjected to a trial displacement field (u) to exactly match the surface of the experimentally observed shape of the DAAM-particle, while minimizing the cost function:
\begin{equation}
    f(u)=E_{el}+\alpha R^2(T; \partial \Omega_t)+\beta E_{pen}(T)
\end{equation}
where $R(T; \partial \Omega t)$ represents the residual cellular forces exerted outside of the cell-target contact region, defined from the phalloidin and immunostaining. The elastic energy ($E_{el}$) penalizes unphysical solutions in which larger forces producing the same shape, while $\beta E_{pen}$ serves as an anti-aliasing term. The weighing parameters, a~(residual traction) and $\beta~$(anti-aliasing), were both set to 1. Spherical harmonic coefficients up to $l_{max}$= 20 were utilized and normal forces were evaluated on a 21 × 41 grid.\\

\textbf{Interaction between the vesicles}---Two vesicles were left to interact through adhesion. The adhesive energy per node between two vesicles is set to $E_{\rm ad}=2k_BT$. If the vertices from different vesicles come within the interaction range (set to $l_{\rm min}$) then this adhesive energy is taken into account. The process is passive when the interaction is solely through the adhesion between them, and there is no active force imparted on the target vesicle by the bigger cell-like vesicle. An active force is imparted by a CMC node in the cell-like vesicle on a vertex in the target vesicle. To calculate the tangential and normal force imparted, we calculate the interaction force first. Let $V_i$ be the vertex of interest in the target vesicle, and $\hat{n}_i$ be its outward normal (computed by taking the average of the normals for all the triangles having the vertex in common). The normal and tangential components of the interaction force due to the sum of active forces acting on this vertex, $\mathbf{F_i}$, are given by,
\begin{eqnarray}
    \mathbf{F_{\perp}}=(\mathbf{F_i}\cdot \hat{n}_i)\hat{n}_i, \nonumber \\
    \mathbf{F_{\parallel}}=\mathbf{F_i}-(\mathbf{F_i}\cdot \hat{n}_i)\hat{n}_i.
\end{eqnarray}
\\

\textbf{Setup for the engulfment simulation}---
In all the simulations of the engulfment process, we construct two vesicles: i) a big cell-like vesicle with 3127 vertices, out of which 150 vertices represent the curved proteins with intrinsic mean curvature $c_0=1~l_{\rm min}^{-1}$, the protein-protein interaction energy is set to $w=1~k_BT$, and ii) a smaller target vesicle made of 847 vertices with no curved proteins. At the initial time, the two vesicles were nearly spherical in shape and placed very near to each other, within the adhesion distance. The bending rigidity for the cell-like vesicle is set to $20~k_BT$ throughout the paper. We varied the bending rigidity of the target vesicle from $20~k_BT$ to a very high value of $2000~k_BT$, which is nearly a rigid sphere.\\

\textbf{Functionalization of hydrogel beads with TDA}---Poly-Aam-co-AAc microparticles (DAAM particles, termed hydrogel beads here) were synthetized as described previously \cite{Vorselen2020_nat_comm}. To functionalize them, 125$\mu$L of 5\% v/v hydrogel beads were washed twice in activation buffer (100 mM MES (Sigma), pH 6.0, 200 mM NaCl) and subsequently incubated for 15 min in 125$\mu$L of activation buffer with 40 mg/mL 1-ethyl-3-(3-dimethylaminopropyl) carbodiimide (EDC, Sigma), 20 mg/mL N-hydroxysuccinimide (NHS, Sigma) and 0.1~\% (v/v) Tween20 (Sigma). Next, the beads were centrifuged (1 min at 5000 g) and quickly resuspended in $125\mu$L of 0.1 mg/mL Tetradecylamine (TDA, Sigma) dissolved in activation buffer. After 1 h incubation, the solution was centrifuged (1 min at 5000 g) to remove excess TDA and then resuspended with $125~\mu$L of blocking buffer (300 mM Tris pH 9.0, 300 mM NaCl, 100 mM of Ethanolamine (Merck-Millipore). After 30 min incubation, functionalized beads were washed 3 times in $250~\mu$L PBS with 0.1\% (v/v) Tween20. Finally, hydrogel beads were resuspended in $125~\mu$L PBS.\\

\textbf{Live cell in vivo microscopy}--- 
Embryos were mounted in 1\% low melting point agarose (UltraPure LMP Agarose, Invitrogen) in Danieau´s solution on a 35 mm glass bottom dish with 14 mm inner diameter of the glass surface (MatTek) and covered with Danieau´s solution. Embryos were imaged at Leica TCS SP8 or SP8 FALCON STED using a HC PL APO CS2 20 x/0.75 NA or HC PL APO CS2 40x/1.30 OIL immersion objectives. A laser excitation of 488 nm with a HyD or PMT detector and 561 nm with a HyD or PMT detector were used. For live imaging of apoptotic clearance and dispersal, z-stacks with a spacing of 2$~\mu$m between z-slices over a total depth of around 30$~\mu$m were acquired with a temporal resolution around 90 seconds. For high-resolution imaging of phagocytic cups 1$~\mu$m z-slices were acquired with a temporal resolution of 40 seconds. Embryos were imaged at 28°C using a H301-K stage top incubator (Okolab) with a UNO-T-H-CO2 controller (Okolab).\\

\textbf{Analysis of phagocytic efficiency of hydrogel beads}
Quantification of phagocytic activity of hydrogel beads was performed using the ‘Multi-point’ tool in FIJI. Phagocytic efficiency was calculated as the ratio of the number of synthetic targets cleared by the embryonic epithelium over the total number of synthetic targets present in a given field of view.\\

\textbf{Analysis of epithelial arm dynamics}---
Tracking of apoptotic cells and synthetic targets was performed by using the ‘MTrackJ’ plugin \cite{meijering2012methods}. A representative number of motile apoptotic or synthetic targets per embryo were tracked over time. The speed was calculated by the distance that targets moved in X-Y-Z directions between consecutive time frames (time lag tlag$~\approx~$90s). The maximum speed corresponds to the maximum instantaneous speed for each track. For the visualization of the spatial target spreading in the embryo in vivo, the x- and y- coordinates of targets were obtained over a time period of a maximum of 30 minutes. Tracks were aligned and centred at the origin.\\

\textbf{\\\\Movie S1: Passive engulfment of the soft vesicle:}---
{The cell-like vesicle with passive ($F=0$) curved protein engulfing the target vesicle. We set the parameters for a cell-like vesicle $N=3127,~\kappa=20~k_BT,~\rho=4.8\%,~c_0=1~l_{\rm min}^{-1},~w=1~k_BT$. The parameters for a target vesicle are set as $N=847,~\kappa=20~k_BT,~\rho=0\%$. The adhesive energy between the cell-like vesicle and the target vesicle is $E_{\rm ad}=2k_BT$.\\}

\textbf{Movie S2: (Experiment) GUV is engulfed by the macrophage}---{Macrophages phagocytose taut GUVs. The macrophage cytosol is labeled by CellTracker Green CMFDA, and the actin is labeled by LifeAct GFP (green). The GUV is opsonized with fluorescent anti-biotin AlexaFluor 647 (magenta). Each frame is 30 seconds.\\}

\textbf{Movie S3: (Experiment) GUV is pushed by the macrophage}---{Macrophages push and then trogocytose low-tension GUVs. The macrophage cytosol is labeled by CellTracker Green CMFDA, and the actin is labeled by LifeAct GFP (green). The GUV is opsonized with fluorescent anti-biotin AlexaFluor 647 (magenta). Each frame is 30 seconds.\\}

\textbf{Movie S4: (Experiment) GUV is bitten by the macrophage}---{Macrophages trogocytose low-tension GUVs. The macrophage cytosol is labeled by CellTracker Green CMFDA, and the actin is labeled by LifeAct GFP (green). The GUV is opsonized with fluorescent anti-biotin AlexaFluor 647 (magenta). Each frame is 30 seconds.\\}

\textbf{Movie S5: (Experiment) Lymphoma cell is engulfment by the macrophage}---{The example of phagocytosis of a lymphoma cell (magenta) by a lifeact-expressing macrophage (green). Time stamp hours:minutes: seconds.\\}

\textbf{Movie S6: (Experiment) Lymphoma cell is pushed by the macrophage}---{The example of a lifeact-expressing macrophage (green) pushing a lymphoma cell (magenta). Time stamp hours:minutes: seconds.\\}

\textbf{Movie S7: (Experiment) Lymphoma cell got bitten by the macrophage}---{The example of trogocytosis/biting of a lymphoma cell (magenta) by a lifeact-expressing macrophage (green). Time stamp hours:minutes: seconds.\\}

\textbf{Movie S8: Active cell-like vesicle engulfing the target}---{The cell-like vesicle with active ($F\neq0$) curved protein engulfing the target vesicle. We set the parameters for a cell-like vesicle $N=3127,~\kappa=20~k_BT,~\rho=4.8\%,~c_0=1~l_{\rm min}^{-1},~w=1~k_BT,~F=2k_BTl_{\rm min}^{-1}$. The parameters for a target vesicle are set as $N=847,~\kappa=20~k_BT,~\rho=0\%$. The target vesicle maintains an internal osmotic pressure $p=10~k_BTl_{\rm min}^{-3}$. The adhesive energy between the cell-like vesicle and the target vesicle is $E_{\rm ad}=2k_BT$.\\}

\textbf{Movie S9: Active cell-like vesicle pushing the target}---{The cell-like vesicle with active ($F\neq0$) curved protein pushing and finally leaves the target vesicle. We set the parameters for a cell-like vesicle $N=3127,~\kappa=20~k_BT,~\rho=4.8\%,~c_0=1~l_{\rm min}^{-1},~w=1~k_BT,~F=2k_BTl_{\rm min}^{-1}$. The parameters for a target vesicle are set as $N=847,~\kappa=20~k_BT,~\rho=0\%$. The target vesicle maintains an internal osmotic pressure $p=0.5~k_BTl_{\rm min}^{-3}$. The adhesive energy between the cell-like vesicle and the target vesicle is $E_{\rm ad}=2k_BT$.\\}

\textbf{Movie S10: Active cell-like vesicle biting the target}---{The cell-like vesicle with active ($F\neq0$) curved protein biting a portion of the target vesicle. We set the parameters for a cell-like vesicle $N=3127,~\kappa=20~k_BT,~\rho=4.8\%,~c_0=1~l_{\rm min}^{-1},~w=1~k_BT,~F=2k_BTl_{\rm min}^{-1}$. The parameters for a target vesicle are set as $N=847,~\kappa=20~k_BT,~\rho=0\%$. The target vesicle maintains an internal osmotic pressure $p=0.1~k_BTl_{\rm min}^{-3}$. The adhesive energy between the cell-like vesicle and the target vesicle is $E_{\rm ad}=2k_BT$.\\}

\textbf{Movie S11: Active cell-like vesicle biting the target instead of pushing, when held a patch}---{The cell-like vesicle with active ($F\neq0$) curved protein biting the target vesicle instead of pushing while held a circular patch by freezing (shown in black) the vertices of the target. We set the parameters for a cell-like vesicle $N=3127,~\kappa=20~k_BT,~\rho=4.8\%,~c_0=1~l_{\rm min}^{-1},~w=1~k_BT,~F=2k_BTl_{\rm min}^{-1}$. The parameters for a target vesicle are set as $N=847,~\kappa=200~k_BT,~\rho=0\%$. The adhesive energy between the cell-like vesicle and the target vesicle  is $E_{\rm ad}=2k_BT$.\\}

\textbf{Movie S12: In-vivo experiment}---{In vivo tracking (yellow lines) of apoptotic targets (red, Bax+ cells co-expressing the plasma membrane marker Lyn-tdTomato, left), 1.3 kPa hydrogel beads (red, middle) and 80 kPa hydrogel beads (red, right) in Lifeact-GFP (cyan) expressing embryos. Embryos were obtained from the Tg(actb1:Lifeact-GFP) line. Time is indicated in h:min:s. Scale bar: 40 $\mu$m.}

\FloatBarrier

\bibliographystyle{abbrv}
\bibliography{sample}